\documentclass[twocolumn,aps,prc,reprint,superscriptaddress,nofootinbib]{revtex4-1}
\usepackage{amsmath}
\usepackage{amssymb}
\usepackage{graphicx}
\usepackage{xcolor}
\usepackage{soul}
\usepackage{float}
\usepackage{hyperref}
\usepackage[capitalise]{cleveref}
\usepackage{epstopdf}
\usepackage{natbib}
\usepackage{multirow}
\newcommand{\beq}{\begin{equation}}
\newcommand{\eeq}{\end{equation}}
\newcommand{\bea}{\begin{eqnarray}}
\newcommand{\eea}{\end{eqnarray}}
\newcommand \hmu {\hat{\mu}}
\newcommand{\avg}[1]{\big\langle #1\big\rangle}

\begin{document}
\title{
Updated Hadron List for Transport Simulations of Heavy-Ion Collisions}


\author{Jordi Salinas San Mart\'in}
\affiliation{Illinois Center for Advanced Studies of the Universe, Department of Physics, University of Illinois at Urbana-Champaign, Urbana, IL 61801, USA}

\author{Renan Hirayama}
\affiliation{Institute for Theoretical Physics, Goethe University, Max-von-Laue-Strasse 1, 60438 Frankfurt am Main, Germany}
\affiliation{Frankfurt Institute for Advanced Studies, Ruth-Moufang-Strasse 1, 60438 Frankfurt am Main, Germany}
\affiliation{Helmholtz Research Academy Hesse for FAIR (HFHF), GSI Helmholtz Center,\\
Campus Frankfurt, Max-von-Laue-Strasse 12, 60438 Frankfurt am Main, Germany}

\author{Jan Hammelmann}
\affiliation{Institute for Theoretical Physics, Goethe University, Max-von-Laue-Strasse 1, 60438 Frankfurt am Main, Germany}
\affiliation{Frankfurt Institute for Advanced Studies, Ruth-Moufang-Strasse 1, 60438 Frankfurt am Main, Germany}

\author{Jamie M. Karthein}
\affiliation{Center for Theoretical Physics -- a Leinweber Institute, Massachusetts Institute of Technology, Cambridge, MA 02139, USA}

\author{Paolo Parotto}
\affiliation{Pennsylvania State University, Department of Physics, University Park, PA 16802, USA}

\author{Jacquelyn Noronha-Hostler}
\affiliation{Illinois Center for Advanced Studies of the Universe, Department of Physics, University of Illinois at Urbana-Champaign, Urbana, IL 61801, USA}

\author{Claudia Ratti}
\affiliation{Physics Department, University of Houston, Houston TX 77204, USA}

\author{Hannah Elfner}
\affiliation{GSI Helmholtzzentrum für Schwerionenforschung, Planckstr. 1, 64291 Darmstadt, Germany}
\affiliation{Institute for Theoretical Physics, Goethe University, Max-von-Laue-Strasse 1, 60438 Frankfurt am Main, Germany}
\affiliation{Frankfurt Institute for Advanced Studies, Ruth-Moufang-Strasse 1, 60438 Frankfurt am Main, Germany}
\affiliation{Helmholtz Research Academy Hesse for FAIR (HFHF), GSI Helmholtz Center,\\
Campus Frankfurt, Max-von-Laue-Strasse 12, 60438 Frankfurt am Main, Germany}


\begin{abstract}
Hadronic transport approaches used in heavy-ion collision simulations rely on a consistent and accurate hadron list with decay channels.
Hadron lists in common use are often experimentally outdated, or, as with the Particle Data Group (PDG) compilations, incompatible with transport codes without further adaptation.
We construct PDG2021+, an updated hadron list including all states from the 2021 Particle Data Booklet, together with a binary-decay list designed for direct use in the SMASH transport framework.
Using the hadron resonance gas model, we validate the PDG2021+ list against lattice quantum chromodynamics results and experimental yield data. 
We show that employing $1\to2$-body decay chains as a proxy for the full decay processes has a suppressing effect in the low-$p_T$ region of the pion spectrum and introduces a $\sim3\%$ systematic uncertainty in the pion $\langle p_T\rangle$.
Moreover, the inclusion of additional states in PDG2021+ further shifts the pion $\langle p_T\rangle$.
These result establish PDG2021+ as a robust, transport-ready hadron list and quantify the systematic effects of decay modeling on key heavy-ion observables.
\end{abstract}
\maketitle

\section{Introduction}
\label{sec:introduction}
Prior to the advent of QCD, particle accelerators were constantly reporting the production of a great number of mostly unstable states -- then referred to as elementary particles on the same footing as electrons or protons. 
Soon after, it was understood that smaller, more fundamental constituents formed these unstable particles, which were re-classified as resonances rather than elementary. 
The observation of an exponentially-rising mass spectrum for these resonances motivated Hagedorn to formulate and introduce the bootstrap model, in which heavy resonances decay into lighter states that can also decay themselves \cite{Hagedorn:1965st, Hagedorn:1968jf}.
A consequence of this idea was that the energy in these systems could be utilized to create heavier mass resonances rather than increase the momentum of the particles, representing an important insight into hadron production. 
Furthermore, the bootstrap model and its extensions were of key importance to develop the thermodynamic description of nuclear matter \cite{Huang:1970iq, Chodos:1974je, Kapusta:1981ay, Frautschi:1971ij, Kapusta:1982qd, Venugopalan:1992hy}. 
This non-kinetic description led to a maximum temperature for hadronic nuclear matter, known as the Hagedorn temperature $T_H$. 
Cabibbo and Parisi postulated that $T_H$ could be indicative of a transition to different degrees of freedom, namely deconfined strongly interacting matter \cite{Cabibbo:1975ig}. 
In fact, the value of $T_H$ predicted by the bootstrap model was found to be remarkably close to the pseudo-critical temperature later determined by lattice QCD calculations \cite{Aoki:2006br}. 

Over the last decades, there has been a collective effort in detecting, identifying, and classifying hadronic resonances, as shown in the Review of Particle Physics \cite{ParticleDataGroup:2022pth}. 
The progress on the experimental side was met with the development of the hadron resonance gas (HRG) model on the theory side, which picks up on the bootstrap model but reduces the number of free parameters while generalizing the model to include all particle species. 
For instance, hadron yield measurements at the LHC indicated a tension between the proton, pion, and strange baryons with thermal model comparisons \cite{Floris:2014pta}, which may be resolved with additional hadronic states \cite{Bazavov:2014xya}. 
Simultaneously, recent lattice QCD results with smaller uncertainties have made it possible to probe the region of the QCD phase diagram near the line of vanishing baryonic chemical potential. 
Furthermore, comparisons between the HRG model and lattice QCD have been shown to be in agreement at temperatures below $T_\text{p.c.}\sim 155$ MeV, adding confidence to the resonance gas description of the hadronic phase of nuclear matter \cite{Chatterjee:2013yga,Bazavov:2014xya,Alba:2017mqu,Chatterjee:2017yhp}. 

Nowadays, hadron lists are typically needed when constructing new equations of state that are later used in hydrodynamical simulations, as well as for freeze-out and sampling codes, and hadronic transport models.
Examples of commonly used hadron lists can be found in \cite{Bass:1998ca, Bleicher:1999xi, Petersen:2008dd, Wheaton:2004qb, Werner:2010aa, Alba:2017mqu, Alba:2020jir} where, in some instances, additional states from the quark model have been included \cite{Capstick:1986ter,Ebert:2009ub}. 
To remain consistent throughout a heavy-ion collision simulation, however, a single hadron list must be used at every step of the evolution.
Moreover, it has been shown that more comprehensive lists that contain all the hadronic states observed by the Particle Data Group (PDG) tend to be more consistent with a number of observables, as opposed to more restrictive sets that only include well-established states \cite{Alba:2017mqu}. 
Recently, the Simulating Many Accelerated Strongly-interacting Hadrons (SMASH) hadronic transport approach has been used to investigate heavy-ion collisions in the low and high energy regimes \cite{SMASH:2016zqf, Weil:2016zrk, Schafer:2021csj} and in the intermediate energy regime to perform studies on baryon stopping \cite{Mohs:2019iee}. 
In this work, we present the PDG2021+ list\footnote{This list is based on the 2021 edition of the Review of Particle Physics \cite{ParticleDataGroup:2020ssz}. By the time of writing, the 2023 edition was just released but it includes no significant updates for our purposes.}, a state-of-the-art hadron list for a consistent use throughout all stages of heavy-ion collision simulations that presents the latest experimental information available from the PDG, including all observed resonances and decays.
For particles and decay channels where experimental information is limited, branching ratios of heavy resonances are modeled through radiative decays into lighter states with matching quantum numbers.

This paper is organized as follows. 
Section \ref{sec:PDG2021Plus} introduces the PDG2021+ particle list, providing comparisons to other commonly used resonance lists. 
Section \ref{sec:validation} presents the effects of using the PDG2021+ extended list on partial pressures, susceptibilities of conserved charges, and particle yields, with specific focus on the strange sector, compared to lattice QCD calculations and experimental data as a validation procedure.
The consequence of incorporating additional resonances by using the PDG2021+ list in the SMASH hadronic framework is analyzed in Sec. \ref{sec:momentum_spectra} in the context of the transverse-momentum spectra of identified particles.
Finally, the conclusions are presented in Sec. \ref{sec:conclusions}.

\section{The PDG2021+ hadron list}
\label{sec:PDG2021Plus}
The Particle Data Group collaboration produces each year a summary of experimental results, including the properties of fundamental particles and hadronic resonances. 
Particle listings are organized according to a confidence level scale, depending on the amount of evidence to back up the existence of each particle and its properties. 
In the case of baryons, the most well-established states are marked with four stars (****), while resonances that have minimal information are given one star (*).
Mesons, on the other hand, are classified as confirmed, unconfirmed, and as ``Further States''.
In this work, we refer to PDG2021 as the collection of reasonably well-established hadrons (baryons with four stars and all confirmed mesons, see \cref{appendix:PDG2021}); in addition, we provide a different version of the list (labeled PDG2021+) with a more relaxed criteria, made up from baryons with at least one star and both confirmed and unconfirmed mesons.
Note, however, that particles under the ``Further States'' section often lack information on quantum numbers, posses large mass uncertainties, and lack decay channel information and thus, are excluded from the list considered here.

In Ref. \cite{Alba:2017mqu}, it was shown that the PDG2016+ list, containing all light and strange hadrons known at the time, was more suitable to adequately reproduce the strange charge susceptibilities and baryon partial pressures calculated using lattice QCD. 
The same list was later used in Refs. \cite{Alba:2017hhe, Alba:2020jir} to demonstrate the effects of additional resonances on particle momentum spectra and the extracted freeze-out parameters from thermal fits, where it was found that missing states led to a better agreement with experimental data and improved yield fits.
Similar studies have also suggested the need for additional states, particularly in the strange sector, to account for differences with other lattice observables \cite{Karsch:2017zzw,Vovchenko:2017uko,Karthein:2021cmb}.
In recent years, experimental observations have indeed determined the existence of previously unknown hadronic states.
On the other hand, along with the newly observed states, a few formerly unconfirmed particles have been removed from the listings as the experimental information has been considered insufficient to support their existence.
A summary of the changes between the 2021 edition of the PDG Particle Review and the 2016 issue is provided in \cref{tab:particle_changes} \cite{Belle:2018mqs, COMPASS:2018uzl, BESIII:2019wkp, Sarantsev:2019xxm, ParticleDataGroup:2022pth}.

\begin{table}[!htbp]
    \begin{tabular}{|c|c|c|}
        \hline
        \multicolumn{3}{|c|}{Particle name}                 \\ \hline
        \multicolumn{2}{|c|}{Added}   &       Removed       \\ \hline
        \hline
        $\pi_2(2005)$   &   $\Sigma(2010)$  &   $a_1(1420)$ \\ \hline
        $X(2370)$       &   $\Sigma(2110)$  &   $X(1840)$   \\ \hline
        $\Lambda(2000)$ &   $\Sigma(2230)$  &   $a_6(2450)$ \\ \hline
        $\Lambda(2070)$ &   $\Sigma(2455)$  &   $\Sigma(1770)$ \\ \hline
        $\Lambda(2080)$ &   $\Sigma(2620)$  &   $\Sigma(1840)$ \\ \hline
        $\Omega(2012)$  &   $\Sigma(3170)$  &   $\Sigma(2000)$ \\ \hline
    \end{tabular}
    \caption{Particle changes on the PDG2021+ hadron list with respect to the previous PDG2016+. It is understood that each particle includes all the elements of the corresponding multiplet and their antiparticles.}
    \label{tab:particle_changes}
\end{table}

\subsection{Characteristics of the list}
The PDG2021+ resonance list\footnote{Available at \url{https://github.com/jordissm/PDG2021Plus}.} is built on the previous PDG2016+, including particles and properties, such as particle identification number (PID), mass, width, degeneracy, baryon number, strangeness content, isospin, electric charge, and branching ratios of (strongly-interacting) decay channels.
However, the new version of the list has some notable differences with respect to its predecessor.

An extensive revision of the PDG2016+ was carried out, updating the values of mass and width, as well as decay channels and branching ratios to the most recent experimental data available. 
For increasingly heavy resonances, information on particle properties such as spin and parity, and branching ratios becomes less certain; in several instances, the reported experimental branching ratios do not add up to 1. 
Although the treatment of decay channels of heavy resonances could in principle affect the final spectra of stable particles, their production is exponentially suppressed.
Previously, for example, the PDG2016+ list assigned a large ($\gtrsim 70 \%$) branching ratio to radiative decays of the form $N_2 \rightarrow N_1 + \gamma$, where $N_2$ and $N_1$ are hadrons with the same baryon number $(B)$, strangeness $(S)$, and electric charge $(Q)$ quantum numbers (here $N_1$ is the next in order of descending mass compatible with such a decay) as a substitute for missing decay channels, while splitting the remaining percentage between available values. 
In contrast, the PDG2021+ list includes the experimentally reported values without modification, only adding radiative decays as a complement to reach the correct normalization. Alternatively, we also provide a version of the decay list where complementary radiative decays are removed; here, decays of heavy resonances for which information is scarce are approximated by the decays of the nearest lighter resonance with known decays and the closest value of $J^P$.
We explicitly verified that including or excluding radiative decay channels in the PDG2021+ list changes the mean transverse momenta of pions, kaons, and protons by less than $0.2\%$, with no appreciable effect on their final spectra.

\begin{figure}[ht]
    \includegraphics[width=0.482\textwidth]{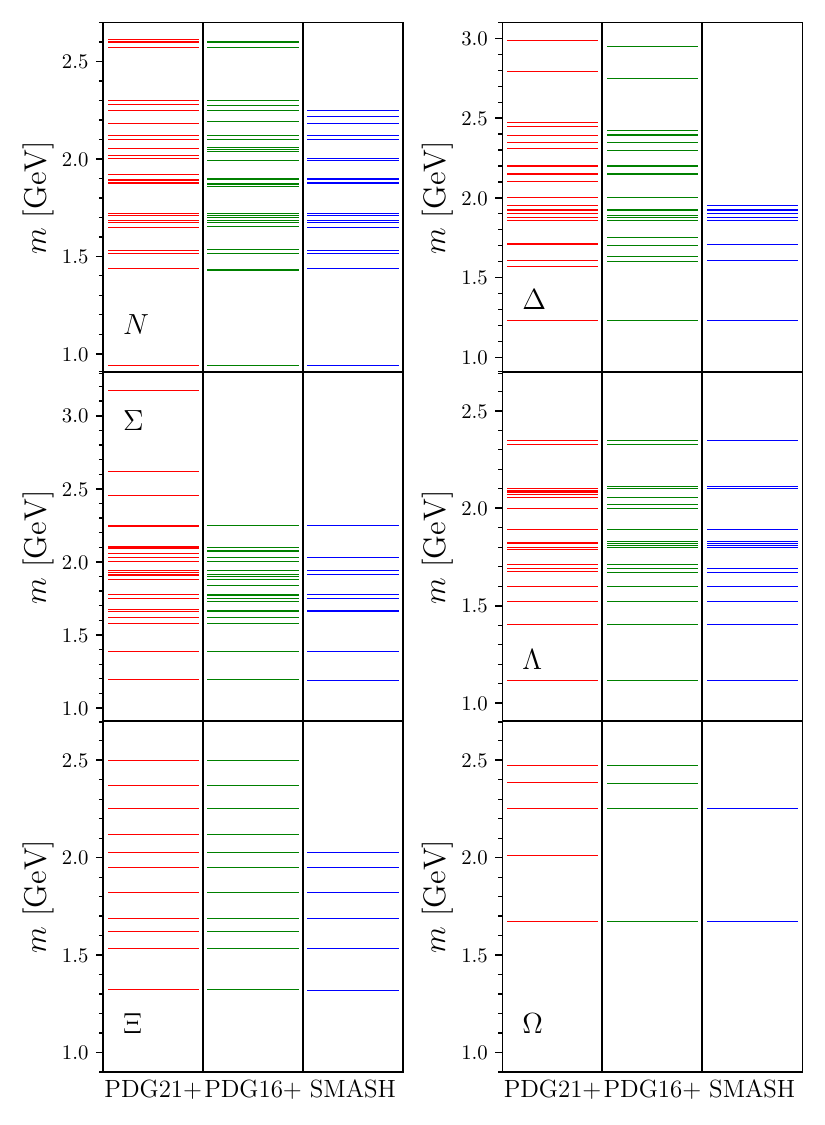}
    \centering
    \caption{Comparison of baryon resonance mass spectra per family between the new PDG2021+ (red), previous PDG2016+ (green), and SMASH (blue) hadron lists. Particle lists based on the latest experimental data available include states that are not completely established and are subject to change in their existence and properties.}
    \label{fig:PDG21Plus_vs_PDG16Plus_vs_SMASH_spectra_comparison}
\end{figure}

The most common approach in thermal and hadronic transport models that take hadronic spectra as input has been to include only the most well-established resonances.
Notably, SMASH uses an extensive set of established hadrons while also allowing for the possibility of including more particles as necessary.
Figure~\ref{fig:PDG21Plus_vs_PDG16Plus_vs_SMASH_spectra_comparison} shows the spectra comparison per baryon family between the PDG2021+, PDG2016+, and SMASH hadron lists\footnote{SMASH version 2.2 was used for all results presented in this paper.}. 
The extended lists contain several resonances that require additional experimental support and are subject to change in properties and quantum numbers.
Most noticeably, these extended lists add in high-mass states for every family when compared to the default SMASH collection of particles.
We also highlight the incorporation of a new, relatively light, $\Omega$ baryon at $m\simeq 2.012$ GeV on the PDG2021+ list with respect to PDG2016+; due to the small number of states in this family, the contribution represents a 25\% increase on the number of triple-strange hadrons.

Alongside with changes on the number of particles and their respective decays, the PDG2021+ list also introduces some changes on the PID of resonances with respect to the standard Monte Carlo numbering scheme \cite{ParticleDataGroup:2022pth}. 
Since SMASH uses the particle identification number to work out the quark content of a baryon, a consistent treatment of the numbering scheme is necessary, to implement the hadron list into the model. 
The details on the modified numbering scheme can be found in Appendix \ref{appendix:PIDs}.

\section{Validation of the PDG2021+ hadron list}
\label{sec:validation}
\subsection{Hadron resonance gas}
The success of the Hagedorn bootstrap idea lies in predicting an exponentially rising spectrum and a maximum temperature for hot hadronic matter, above which a phase transition may occur.
In order to extract such maximum temperature, known as Hagedorn temperature, one can perform a fit of the hadronic spectrum to the exponentially rising density of states $\rho$ of the bootstrap model.
This procedure serves as a quantitative validation of hadron lists if the extracted Hagedorn temperature approximates the pseudo-critical temperature $T_\text{p.c.}$ from lattice QCD.
As shown in \cref{appendix:hadron_spectrum}, the limiting temperature extracted from the PDG2021+ is indeed closer to $T_\text{p.c.}$ than that from the SMASH list of hadrons.
An alternative approach to this continuous spectrum is to use the experimentally available discrete hadronic states in replacement of the density of states of \cref{eq:bootstrap_solution}. 
In practice, this method is known as the hadron resonance gas model and is useful to compute the thermodynamics of a system of non-interacting hadrons and their resonances.
While this approach reduced the theoretical uncertainty, as it makes direct connection to experimental information, the uncertainty on missing resonances is still debated \cite{Broniowski:2000bj,Chatterjee:2009km,Noronha-Hostler:2016ghw}.

The HRG model is a widely-used theoretical framework for describing the properties of nuclear matter at finite density and low temperature ($T\lesssim 155$ MeV). 
The fundamental assumption of the model is that the system is well described by an uncorrelated gas of hadrons and resonances, with the bulk thermodynamics of QCD calculated within the grand-canonical ensemble. 
The hadrons and resonances are thus treated as point-like particles whose properties are determined by their quantum numbers, such as mass, spin, and isospin. 

Validation for the HRG model comes from the remarkable agreement with predictions from lattice calculations \cite{Borsanyi:2014ewa,Bellwied:2019pxh} and has been successfully applied to describe further observables, such as particle yields \cite{Andronic:2017pug} and net-particle fluctuations \cite{Alba:2014eba,Bellwied:2018tkc,Bluhm:2018aei,Alba:2020jir}. 
However, because of its assumptions, the model has some limitations when applied to the hadron phase of a heavy-ion collision which include the finite-size of constituents, the non-zero width of broad resonances, and interactions between hadrons. 
To address these limitations, some extensions to the HRG model have been proposed such as the inclusion of an excluded-volume \cite{Yen:1997rv, Andronic:2012ut, Noronha-Hostler:2012ycm, Bhattacharyya:2013oya, Vovchenko:2014pka, Albright:2015uua, Satarov:2016peb, Vovchenko:2017xad, Alba:2017bbr, Vovchenko:2018eod, Motornenko:2020yme, Karthein:2021cmb}, van der Waals terms \cite{Vovchenko:2016rkn, Vovchenko:2017zpj, Vovchenko:2017uko, Samanta:2017yhh, Sarkar:2018mbk, Vovchenko:2020lju}, mean-field interactions \cite{Huovinen:2017ogf, Steinert:2018zni, Sorensen:2020ygf}, quantum effects under the Beth-Uhlenbeck approach \cite{Vovchenko:2017drx,Hanafy:2021ffj}, or based on scattering phase shifts \cite{Dashen:1969ep,Venugopalan:1992hy,Friman:2015zua,Lo:2017lym,Fernandez-Ramirez:2018vzu,Dash:2018can,Dash:2018mep,Dash:2019zzo}.
Despite the improved descriptions that these extensions may provide, in this work we use the ideal HRG for its clarity to demonstrate the effect of adding more resonances.

In the ideal HRG model, the total pressure of the gas is obtained by summing over all individual pressures of the spectrum, 
\begin{equation}\label{eq:total_pressure}
    \frac{p}{T^4} = \sum_i\frac{p_{B_iS_iQ_i}}{T^4},
\end{equation}
which are given by
\begin{widetext}
\begin{align}\label{eq:individual_pressure}
    \frac{p_{B_iS_iQ_i}}{T^4} =&\, (-1)^{B_i+1}\frac{1}{T^3}\frac{d_i}{2\pi^2}\int_0^\infty dp\ p^2 \ln\Bigg[1 + (-1)^{B_i+1}\exp\left( {-\frac{\sqrt{p^2+m_i^2}}{T}-\hmu_i} \right) \Bigg],
\end{align}
\end{widetext}
where $\hmu_i = \sum_q X_i^q \mu_q/T$ is the dimensionless total chemical potential, $\mu_q$ is the chemical potential for the conserved charge $q$, and the index $i$ runs over all particles. 
Each distinct particle possesses a spin degeneracy $d_i$, mass $m_i$, and quantum numbers $X_i^q=\{ B_i,S_i,Q_i \}$. 
In addition, mesons and baryons are described by different statistics, namely, Bose-Einstein for the former and Fermi-Dirac for the latter. 
Notice that in both cases the contribution is positive, signaling an increase of pressure for every additional state considered.

The number density for particle species $i$ can be found from the total pressure in \cref{eq:total_pressure} with the help of the thermodynamic relation
\begin{equation}\label{eq:number_densities}
    n_{i}(T,\mu_B,\mu_S,\mu_Q) = \frac{\avg{ N_i^\text{prim} }}{V} = \frac{1}{T}\frac{\partial p}{\partial \hmu_i}\Bigg\lvert_{T},
\end{equation}
where $\avg{ N_i^\text{prim} }$ is the (primordial) mean multiplicity for species $i$ in a volume $V$. 
In reality, the predicted yield of a particle is given by the sum of the primordial multiplicity and the contribution from all resonances that decay into a final state containing such particle, i.e.,
\begin{equation}\label{eq:particle_multiplicity}
    \avg{ N_i }
    = n_{i}(T,\mu_B,\mu_S,\mu_Q)\ V 
      + \sum_R \Gamma_{R\rightarrow i} \avg{ N_R^\text{prim} },
\end{equation}
where $\avg{ N_R^\text{prim} }$ is the mean multiplicity of resonance $R$ and $\Gamma_{R\rightarrow i}=\sum_\text{chan.}g_i\ \gamma_{R\rightarrow i}$ is the sum across all decay channels of resonance $R$ with corresponding branching ratios $\gamma_{R\rightarrow i}$, that have $g_i$ occurrences of particle $i$ as a decay product.
\subsection{Comparisons to lattice QCD}
\subsubsection{Partial pressures}
Within the HRG model, the pressure at finite chemical potential can be obtained by integrating \cref{eq:individual_pressure}. After a low-temperature expansion, detailed in \cref{subsubsec:partial_pressures}, the total pressure becomes
\begin{align}
    \frac{p}{T^4} &= \tilde{\phi}_{0}(T) + \sum_{B,S} \tilde{\phi}_{BS}(T)\, \cosh(B\hmu_B+S\hmu_S) \nonumber \\
    &= \tilde{\phi}_{0}(T) 
    + \tilde{\phi}_{0|1|}(T)\cosh(\hmu_S) \nonumber \\
    & \hspace{39pt} + \tilde{\phi}_{10}(T)\cosh(\hmu_B) \nonumber \\
    & \hspace{39pt} + \tilde{\phi}_{1|1|}(T)\cosh(\hmu_B-\hmu_S) \nonumber \\
    & \hspace{39pt} + \tilde{\phi}_{1|2|}(T)\cosh(\hmu_B-2\hmu_S) \nonumber \\
    & \hspace{39pt} + \tilde{\phi}_{1|3|}(T)\cosh(\hmu_B-3\hmu_S),
\label{eq:HRG_pressure_sector_decomposition}
\end{align}
where each term in the sum corresponds to the partial pressure associated with a particular set of quantum numbers. Notice that the dimensionless pressure coefficients $\tilde{\phi}_{BS}$ are now linear combinations of the more general $\phi_{BSQ}$; for instance, the kaon contribution is $\tilde{\phi}_{0|1|}=\phi_{0|1|0}+\phi_{0|1||1|}$.
For further details see \cite{Bazavov:2013dta, Noronha-Hostler:2016rpd, Alba:2017mqu}.

\begin{figure}[!b]
    \includegraphics[width=0.95\linewidth]{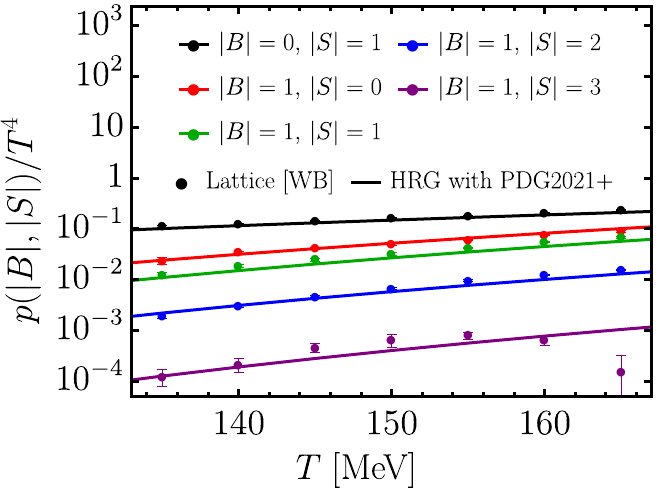}
    \centering
    \caption{Contribution to the total pressure of strange mesons and different baryonic sectors within the ideal hadron resonance gas using the PDG2021+ particle list, compared to the lattice results from \cite{Alba:2017mqu}.}
    \label{fig:all_partial_pressures}
\end{figure}

In \cref{fig:all_partial_pressures}, we show the contribution of each of the partial pressures to the total pressure.
Results from the HRG model calculations, as described above, with the updated PDG2021+ list are compared to the lattice QCD results obtained from imaginary strangeness chemical potential simulations from \cite{Alba:2017mqu}. 
This logarithmic plot illustrates the range in order of magnitude that the different sectors span with smaller contributions from more rare species.
We can see that the agreement achieved between lattice QCD and HRG with the PDG2021+ hadronic list is remarkable across the orders of magnitude covered here.

\begin{figure}[tb]
    \includegraphics[width=0.95\linewidth]{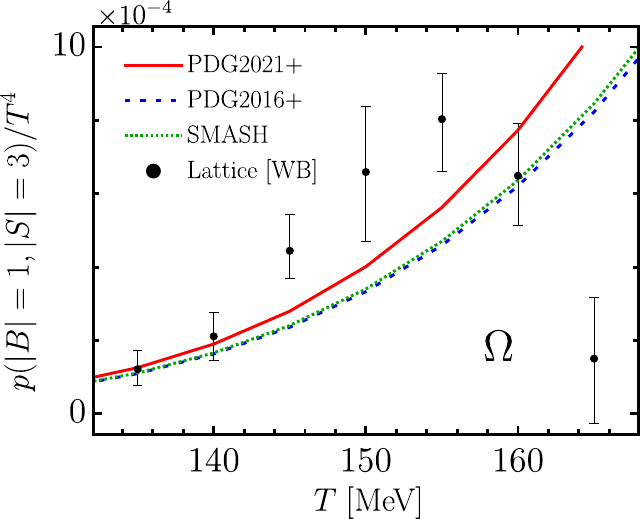}
    \centering
    \caption{Contribution to the total pressure from $\Omega$ resonances within the ideal hadron resonance gas using the PDG2021+ (solid red) particle list, compared to the lattice results from \cite{Alba:2017mqu} and to the PDG2016+ (dashed blue) and SMASH (dotted green) hadron lists.}
    \label{fig:omega_partial_pressures}
\end{figure}

In particular, the triply strange baryons in \cref{fig:omega_partial_pressures} have a better agreement between HRG and lattice with the updated PDG2021+ list, when compared to SMASH and PDG2016+. 
The strangeness $S=3$ baryons are more than 20$\%$ higher at $T\sim150$ MeV for the PDG2021+ list, which is due to the additional $\Omega$ baryon listed by the PDG in 2021. 

\subsubsection{Susceptibilities}
Despite the overall agreement that the HRG model has with results from lattice QCD at temperatures below $T_\text{p.c.}$ for the partial pressures, recent studies have pointed out discrepancies in other observables sensitive to the flavor content of hadrons \cite{Alba:2017mqu,Bellwied:2019pxh}. 
Including more hadrons from a particular sector can improve or amplify the tension when comparing the HRG model against lattice results. 
Examples of such observables include the diagonal and off-diagonal susceptibilities of conserved charges and their ratios, which are calculated according to

\begin{equation}
    \chi_{ijk}^{BSQ} = \frac{\partial^{i+j+k} (p/T^4)}{\partial(\mu_B/T)^i \partial(\mu_S/T)^j \partial(\mu_Q/T)^k}\Bigg\lvert_{\vec{\mu}=0},
    \label{eq:susceptibilities}
\end{equation}
where $i$, $j$, and $k$ are the orders of the susceptibilities corresponding to the $B$, $S$, and $Q$ conserved charges, respectively.

\begin{figure}[!htbp]
    \includegraphics[width=0.95\linewidth]{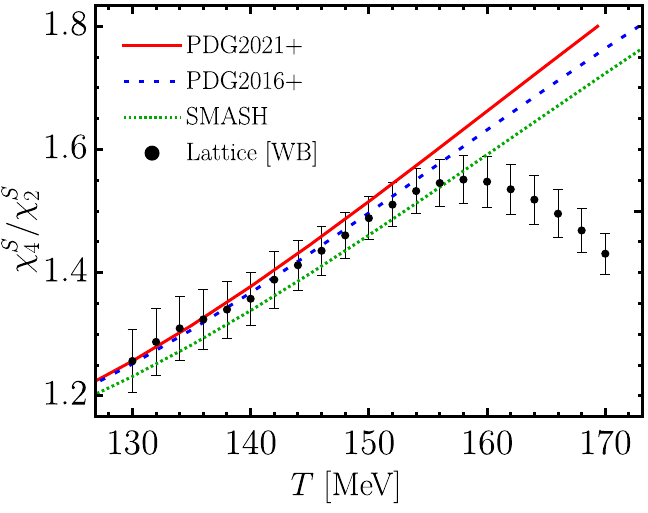}
    \centering
    \caption{Strange susceptibility $\chi_4^S$ at vanishing chemical potential, normalized by $\chi_2^S$ as a function of temperature within the ideal HRG model using the PDG2021+ (solid red), PDG2016+ (dashed blue), and SMASH (dotted green) particle lists, compared to the lattice results from Ref. \cite{Alba:2017mqu}.
    }
    \label{fig:ratio_chi4S_to_chi2S}
\end{figure}

We compare the conserved charge susceptibilities from lattice QCD and ideal HRG calculations with the different hadron lists: SMASH, PDG2016+, and PDG2021+.
It has been previously shown that the excluded-volume dependence can be eliminated when particular second-order diagonal and off-diagonal susceptibility ratios are considered \cite{Karthein:2021cmb}.
This cancellation makes some of these ratios (e.g., $\chi_{11}^{BQ}/\chi_2^B$ and $\chi_{11}^{BS}/\chi_2^B$) safer to be studied using the simpler ideal HRG used here.
Other susceptibility ratios, such as $\chi_{4}^{B}/\chi_2^B$ and $\chi_{4}^{S}/\chi_2^S$, possess no such exact cancellations but the excluded-volume effects are most prevalent for $T\gtrsim T_\text{p.c.}$ and can be assumed to be small at lower temperatures. 
In Fig. \ref{fig:ratio_chi4S_to_chi2S}, the higher order strangeness susceptibility ratio $\chi_4^S/\chi_2^S$ is shown for the different lists. As was shown in Ref.~\cite{Alba:2017mqu}, this particular susceptibility ratio is proportional to the square of the strangeness content $\sim \left\langle S^2 \right\rangle$.
The results indicate that the new list PDG2021+ behaves very similarly to the PDG2016+ list, with differences only at the percent level.
At higher temperatures, these two lists begin to deviate, due to the presence of the new $\Omega$ baryon included in the PDG2021+ list. The SMASH list is consistently below the other two, because it contains fewer multi-strange baryons. 
In general, the agreement with lattice results is very similar for the three lists, owing to the error bars on the lattice QCD data for this observable. 

\subsection{Thermal yield fits to experimental data}
The hadron resonance gas model is also often used to fit the hadron yield data from relativistic heavy-ion collisions, assuming thermal and chemical equilibrium between all stable hadrons and resonances \cite{Andronic:2005yp,Becattini:2012xb,Alba:2014eba,Andronic:2017pug,Andronic:2020iyg,Flor:2020fdw,Flor:2021olm}.
The HRG model fits are performed by minimizing the value of the $\chi^2$ function, 
\begin{equation}
\label{eq:ThermalFit}
    \frac{\chi^2}{N_\text{d.o.f.}} = \frac{1}{N_\text{d.o.f.}} \sum_{i=1}^N \frac{(N_i^{\mathrm{exp}} - N_i^{\mathrm{HRG}})^2}{\sigma_i^2},
\end{equation}
determined from the comparison between experimental yields and yields calculated in the thermal model.
From this $\chi^2$-minimization procedure, the best fitting chemical freeze-out parameters, $T_\text{ch}, \mu_{B,\text{ch}}, V_\text{ch}$ are extracted.

We utilized the Thermal-FIST package to perform the fits \cite{Vovchenko:2019pjl}.
Following the previous results of Refs.~\cite{Bellwied:2013cta,Noronha-Hostler:2016rpd,Bellwied:2018tkc,Bellwied:2019pxh,Flor:2020fdw,Alba:2020jir}, in this manuscript we considered two separate scenarios for chemical freeze-out, namely a single freeze-out and two freeze-out scenario.
In the single freeze-out scenario, a global fit of all hadrons is performed, yielding a single set of $\{T_\text{ch}, \mu_{B,\text{ch}}, V_\text{ch}\}$.
On the other hand, in the two freeze-out scenario we consider the flavor-dependent freeze-out hypothesis following the work in \cite{Flor:2020fdw}. Here, we allow for two sets of freeze-out parameters by fitting light ($\pi,K,p$) and strange ($K,\Lambda,\Sigma,\Xi,\Omega,K^0_S,K^{*0},\phi$) particles separately.

\begin{widetext}
\begin{table*}[!htbp]
    \begin{tabular}{lccccc}
        \toprule
        Particle list & \multicolumn{1}{c}{\hspace{7mm}Particle subset}\hspace{7mm} &
            \multicolumn{1}{c}{\hspace{5mm}$T_\text{ch}$ [MeV]}\hspace{5mm}     &
            \multicolumn{1}{c}{\hspace{5mm}$\mu_{B,\text{ch}}$ (fixed) [MeV]}\hspace{5mm} &
            \multicolumn{1}{c}{\hspace{5mm}$V_\text{ch}$ [fm$^3$]}\hspace{5mm} &
            \multicolumn{1}{c}{\hspace{5mm}$\chi^2/N_\text{d.o.f.}$}\hspace{5mm} \\ \hline
        \multirow{3}{*}{SMASH list} 
        & \multicolumn{1}{c}{All}  & \multicolumn{1}{c}{$157.65\pm 2.24$} & \multicolumn{1}{c}{$0$} & \multicolumn{1}{c}{$5075.0\pm603.7$} & \multicolumn{1}{c}{$78.20/7$} \\ 
        & \multicolumn{1}{c}{Light}  & \multicolumn{1}{c}{$146.58\pm 2.69$} & \multicolumn{1}{c}{$0$} & \multicolumn{1}{c}{$8631.94\pm1239.49$} & \multicolumn{1}{c}{$5.07/1$} \\
        & \multicolumn{1}{c}{Strange}  & \multicolumn{1}{c}{$169.43\pm 2.84$} & \multicolumn{1}{c}{$0$} & \multicolumn{1}{c}{$3030.02\pm417.28$} & \multicolumn{1}{c}{$40.42/5$} \\
        \hline
        \multirow{3}{*}{\shortstack{PDG2021+ list\\ ($1 \to 2$ decays)}}
        & \multicolumn{1}{c}{All}  & \multicolumn{1}{c}{$154.80\pm 2.22$} & \multicolumn{1}{c}{$0$} & \multicolumn{1}{c}{$5377.59\pm674.72$} & \multicolumn{1}{c}{$28.96/7$} \\ 
        & \multicolumn{1}{c}{Light}  & \multicolumn{1}{c}{$145.57\pm 2.66$} & \multicolumn{1}{c}{$0$} & \multicolumn{1}{c}{$8628.31\pm1269.67$} & \multicolumn{1}{c}{$5.18/1$} \\ 
        & \multicolumn{1}{c}{Strange}  & \multicolumn{1}{c}{$163.04\pm2.13$} & \multicolumn{1}{c}{$0$} & \multicolumn{1}{c}{$3667.39\pm409.82$} & \multicolumn{1}{c}{$3.03/5$} \\ 
        \hline
        \multirow{3}{*}{\shortstack{PDG2021+ list\\ ($1 \to \text{all}$ decays)}}
        & \multicolumn{1}{c}{All}  & \multicolumn{1}{c}{$155.01\pm2.15$} & \multicolumn{1}{c}{$0$} & \multicolumn{1}{c}{$5287.69\pm639.26$} & \multicolumn{1}{c}{$26.41/7$} \\ 
        & \multicolumn{1}{c}{Light}  & \multicolumn{1}{c}{$146.96\pm2.67$} & \multicolumn{1}{c}{$0$} & \multicolumn{1}{c}{$7717.14\pm1139.73$} & \multicolumn{1}{c}{$0.61/1$} \\ 
        & \multicolumn{1}{c}{Strange}  & \multicolumn{1}{c}{$162.59\pm1.42$} & \multicolumn{1}{c}{$0$} & \multicolumn{1}{c}{$3753.62\pm283.55$} & \multicolumn{1}{c}{$2.50/5$} \\ 
        \toprule
    \end{tabular}
    \caption{Summary from thermal yield fits (see \cref{fig:yield_ALICE_PDG21Plus,fig:yield_ALICE_SMASH}) performed for the SMASH and PDG2021+ hadron lists, including chemical freeze-out parameters ($T_\text{ch}$, $\mu_{B,\text{ch}}$, and $V_\text{ch}$) and fit quality $\chi^2/N_\text{d.o.f.}$. 
    The ``All'' results correspond to the single freeze-out scenario, where all particles are fitted simultaneously while the ``Light'' and ``Strange'' results correspond to the two freeze-out scenario, in which a separate fit is performed for both sets of particles (see main text for details). 
    In all cases, the PDG2021+ is in better agreement with the experimental data.  Experimental data points correspond to ALICE Pb--Pb $0$–$10\%$ collisions at $5.02$ TeV \cite{ALICE:2019hno,ALICE:2021ptz,Bhasin:2022rpo}. 
    }
    \label{tab:fit_summary}
\end{table*}
\end{widetext}
The thermal yield fits for both the PDG2021+ and SMASH lists are provided in \cref{appendix:thermal_models} in \cref{fig:yield_ALICE_PDG21Plus,fig:yield_ALICE_SMASH}.
A summary of the fit results is shown in \cref{tab:fit_summary}.
The fit quality, indicated by the $\chi^2/N_\text{d.o.f.}$ value, is consistently superior when using the more comprehensive PDG2021+ list for all particle species in both freeze-out scenarios. 
Notably, the correlation between thermal yields and experimental values in the strange sector sees significant improvement, with the sole exception of $\Lambda$ particles. 

\section{Transverse-momentum spectra and \texorpdfstring{$\avg{p_T}$}{<p\_T>}}
\label{sec:momentum_spectra}
Previous studies  have found that the inclusion of extra hadronic states can influence both the spectra and mean transverse momentum $\langle p_T\rangle$ \cite{Alba:2017hhe} as well as the flow harmonics \cite{Noronha-Hostler:2013ria,Alba:2017hhe}.
To thoroughly check the influence of these new states in heavy-ion collision simulations there are a number of steps that would need to be taken.
First, we would need to include these extra resonances in the hadron resonance gas equation of state and match that equation of state to lattice QCD results.
Then, we would need to run full hydrodynamic simulations with this new equation of state and new resonances, retuning parameters such as the normalization constant, shear and bulk viscosities, etc.
Such study was done previously in \cite{Alba:2017hhe} for the PDG2016+ list but only considering the influence of new states on the direct decays.
A more in depth study is still needed where the full effect of hadronic rescattering and the interplay with these new states is taken into account.
While we plan to study this in detail in a future work, in this paper we take a simpler approach and instead incorporate these new states in a blast-wave model where we consider both the possibility of their influence on just direct decays as well as their influence on hadron rescattering using the updated SMASH code.

The blast-wave model is a simplified, parametric approach that has been often used to describe the spectra and elliptic flow observed in relativistic heavy-ion collisions \cite{Siemens:1978pb, Huovinen:2001cy, STAR:2001ksn, Rybczynski:2012ed, Jaiswal:2015saa, ALICE:2019hno, Vovchenko:2020kwg, Chen:2020zuw}. In its simplest form, the model assumes a simultaneous freeze-out of all particles on a hypersurface in local thermal equilibrium with boost-invariance for the longitudinal expansion, and a simple parametrization for the transverse flow,
\begin{equation}
    \beta_T(r_\perp) = \beta_s\left( \frac{r_\perp}{R} \right)^n,
\end{equation}
where $\beta_s$ is the maximum surface flow velocity, $r_\perp=(r_x^2+r_y^2)^{1/2}$ is the transverse radius, $R$ is the radius of the expanding medium, and $n$ changes the flow profile. In the blast-wave model, the particle momentum spectra is 
\begin{align}
    \frac{dN}{p_Tdp_T} \sim \int_0^R dr\,r\,m_T\,I_0\left(\frac{p_T\sinh\rho}{T_\text{kin}}\right)K_1\left(\frac{m_T\cosh\rho}{T_\text{kin}}\right),
\end{align}
where $\rho(r_\perp)=\tanh^{-1}\beta_T(r_\perp)$, $K_1$ and $I_0$ are the modified Bessel functions, $m_T=\sqrt{p_T^2+m^2}$ is the transverse mass, and $T_\text{kin}$ is the kinetic freeze-out temperature.

The kinetic freeze-out temperature $T_\text{kin}$, average transverse flow velocity $\avg{\beta_T}$, and parameter $n$ that controls the proportionality of $\avg{\beta_T}$ to the transverse flow velocity at the surface used are those listed in Table 4 of Ref.~\cite{ALICE:2019hno}.
On the other hand particle multiplicities are obtained by applying the Cooper-Frye prescription to sample the multiplicity distribution.
In this procedure, the chemical freeze-out temperature $T_\text{ch}$ and baryon chemical potential $\mu_{B}$  are those resulting from the fits in Sec.~\ref{appendix:thermal_models}.
Alternatively, one could follow the work in Ref.~\cite{Mazeliauskas:2019ifr}, where a fit to the identified particle spectra accounting for the resonance feed-down allows to extract the blast-wave parameters.

\begin{figure}[bp]
    \includegraphics[width=0.48\textwidth]{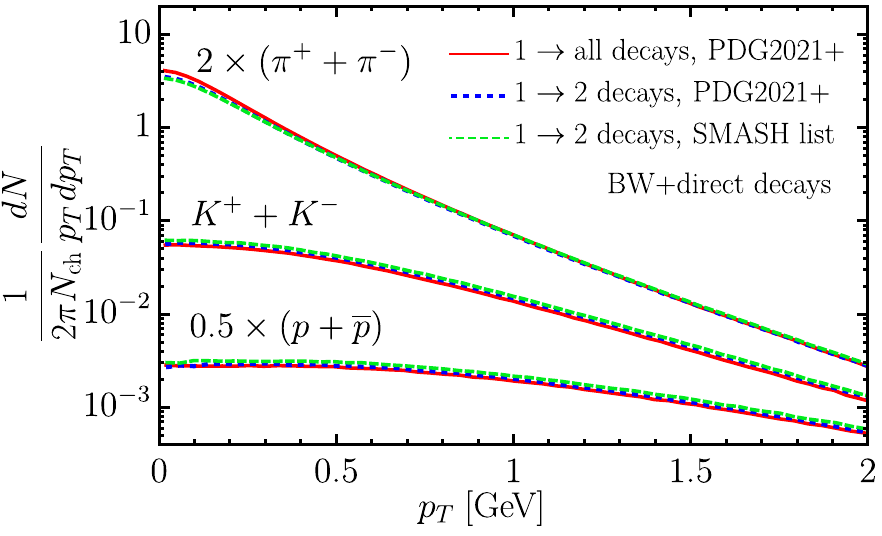}
    \centering
    \caption{Transverse-momentum spectra for $\pi^\pm$, $K^\pm$, and $p(\overline{p})$ for the blast-wave model after direct decays using the PDG2021+ with all decays (solid red), PDG2021+ with intermediate states (dashed blue) and default SMASH (dashed green) hadron lists.}
    \label{fig:direct_decays_momentum_spectra}
\end{figure}

In transport approaches, a collision criterion is needed to determine whether particles collide; the geometric or Bertsch criterion has commonly been used and is based on the distance at closest approach and the geometric interpretation of the cross-section in binary scatterings. A limitation arising from this type of condition is that there is no generalization to deal with multi-particle interactions while preserving detailed balance. Possible solutions to this problem can be achieved using stochastic methods or by using decay chains of $1\rightarrow2$ processes \cite{Bass:1998ca,Cugnon:1980zz,Weil:2016fxr,Staudenmaier:2021lrg}. Here, we have adopted the latter method to include all decays from the PDG2021+ into SMASH. Examples of this approach can already be found in UrQMD and SMASH. 

\begin{figure}[b]
    \includegraphics[width=0.48\textwidth]{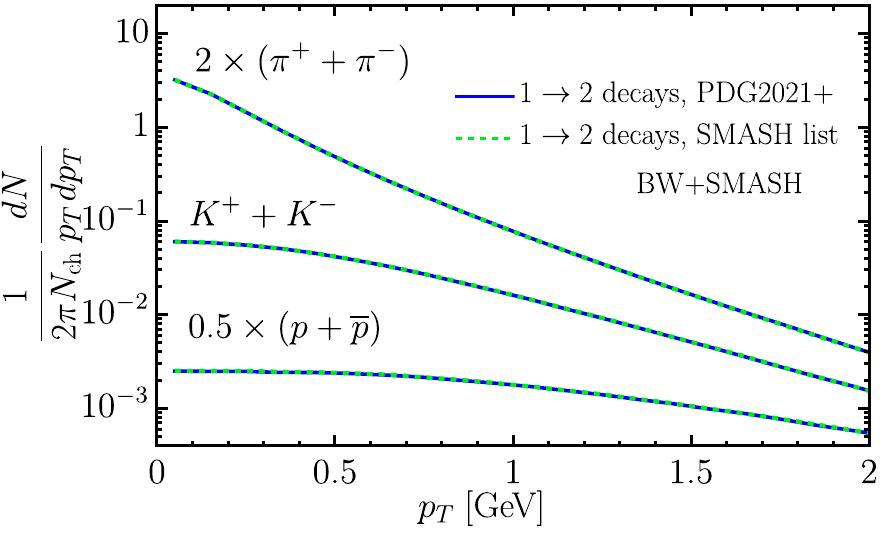}
    \centering
    \caption{Transverse-momentum spectra for $\pi^\pm$, $K^\pm$, and $p(\overline{p})$ for the blast-wave model after the hadronic rescattering phase with SMASH using the PDG2021+ with intermediate states (solid blue) and default SMASH (dashed green) hadron lists.
    }
    \label{fig:SMASH_decays_momentum_spectra}
\end{figure}

In \cref{fig:direct_decays_momentum_spectra} we show the spectra of $\pi^\pm$, $K^\pm$, and $p(\overline{p})$ for Pb--Pb collisions at $\sqrt{s_{NN}}=5.02$ TeV and 0–10$\%$ centrality, as obtained from the cylindrically-symmetric blast-wave model \cite{Schnedermann:1993ws} for the SMASH and PDG2021+ hadron lists with direct decays. For the case of the PDG2021+ list, we include both the cases where multi-body and $1\to 2$ decays are allowed. The effect of introducing intermediate states to account for multi-body processes can be directly seen by comparing the two scenarios that use the PDG2021+ hadronic list. Here, the pion spectra is enhanced at low $p_T$ followed by a more pronounced drop in the case where all decay products are considered. For kaons and protons, the production is consistently higher when using decay chains across the range of $p_T$ examined. This can be explained by the fact that intermediate states possess decay channels other than the desired final state particles, thus affecting the total production of all particle species. Similarly, using a more restrictive set of particles, as in the case of the SMASH list, has the effect of suppressing the number of pions in the low-momentum region while retaining the same slope for $p_T\gtrsim 0.5$ GeV. 

To study the effect of including more hadronic states in a more realistic setting, we show in \cref{fig:SMASH_decays_momentum_spectra} the spectra of $\pi^\pm$, $K^\pm$, and $p(\overline{p})$ for Pb--Pb collisions at $\sqrt{s_{NN}}=5.02$ TeV and 0–10$\%$ centrality for the SMASH and PDG2021+ hadron lists after the hadronic re-scattering phase performed with SMASH. In contrast to the case of direct decays, including more resonances seems to have little effect on all transverse-momentum profiles.

It is worth noting that in all cases we normalize the spectra $\frac{1}{2\pi}\frac{dN}{p_Tdp_T}$ by the total number of charged particles, $N_\text{ch}$. Because we wanted to use identical blast-wave models for each of our lists, there are some minor differences in the total $N_\text{ch}$.  However, in full relativistic viscous hydrodynamics simulations there is an overall normalization factor that provides the scale of initial conditions that is tuned to reproduce $N_\text{ch}$.  Thus, minor differences between $N_\text{ch}$ would be removed from full realistic simulations and, therefore, we normalize by $N_\text{ch}$ to remove any minor deviations there and focus only on the slope of the spectra. 

Finally, in \cref{tab:mean_pT} we show the mean transverse-momentum for $\pi^\pm$, $K^\pm$, and $p(\overline{p})$ as extracted from the blast-wave model with direct decays (BW+DD) and a hadronic re-scattering phase (BW+SMASH) with a comparison to the experimental values for Pb--Pb collisions at $\sqrt{s_{NN}}=5.02$ TeV and 0–10$\%$ centrality. These calculations allow us to test the effect of considering more hadronic states. We found that using an extended hadronic list has a slight but minimal ($\sim 1\%$) effect on the pion mean transverse-momentum in both the BW+DD and BW+SMASH scenarios. For kaons and protons, the results are nearly identical across the lists. 

In the case of BW+DD, we can also test the appropriateness of using $1\to2$-body decay chains as a proxy to the full (true) decay processes. 
We found that using decay chains led to a $\sim 3\%$ overestimation of the pion mean transverse-momentum. 
Meanwhile, the kaon and proton $\langle p_T\rangle$ remain nearly the same. 
Note that this computation is not possible in the full hadronic transport approach because of the limitations introduced in SMASH from detailed balance when using the geometric collision criterion.
However, the effect on the $p(\bar{p})$ spectra from decay chains in the dynamical picture has been explored by the authors of Ref.~\cite{Garcia-Montero:2021haa} and found to be a non-negligible contribution to the proton yield, where the stochastic collision criterion was used to maintain detailed balance.

At this point, we have only tested these results with a blast wave model but it will be interesting to study these effects with full hydrodynamic simulations down the road. 

\begin{widetext}
\begin{table*}[htb]
    \begin{tabular}{lccc|}
        \toprule
        Blast Wave+direct decays (BW+DD)& \multicolumn{1}{p{4.0cm}}{\hspace{14.0mm}$\pi^++\pi^-$} &
              \multicolumn{1}{p{4.0cm}}{\hspace{13.0mm}$K^++K^-$}     &
              \multicolumn{1}{p{4.0cm}}{\hspace{16.5mm}$p+\overline{p}$} \\ \hline
        \multicolumn{1}{l}{SMASH list}                       & \multicolumn{1}{c}{$0.5240\pm 0.0005$}  & \multicolumn{1}{c}{$0.9055\pm 0.0015$} & \multicolumn{1}{c}{$1.4095\pm 0.0025$} \\ 
        \multicolumn{1}{l}{PDG2021+ list ($1\rightarrow$ 2 decays)}    & \multicolumn{1}{c}{$0.5170\pm 0.0005$}  & \multicolumn{1}{c}{$0.9060\pm 0.0015$} & \multicolumn{1}{c}{$1.4110\pm 0.0025$} \\ 
        \multicolumn{1}{l}{PDG2021+ list ($1\rightarrow$ all decays)} & \multicolumn{1}{c}{$0.5020\pm 0.0005$}  & \multicolumn{1}{c}{$0.9050\pm 0.0015$} & \multicolumn{1}{c}{$1.4100\pm 0.0025$} \\ 
        \toprule
        Blast Wave+SMASH (BW+SMASH)   & \multicolumn{1}{p{4.0cm}}{\hspace{14.0mm}$\pi^++\pi^-$} &
              \multicolumn{1}{p{4.0cm}}{\hspace{13.0mm}$K^++K^-$}     &
              \multicolumn{1}{p{4.0cm}}{\hspace{16.5mm}$p+\overline{p}$} \\ \hline
        \multicolumn{1}{l}{SMASH list}                       & \multicolumn{1}{c}{$0.5444 \pm 0.0001$}  & \multicolumn{1}{c}{$0.9390 \pm 0.0003$} & \multicolumn{1}{c}{$1.4931 \pm 0.0008$} \\ 
        \multicolumn{1}{l}{PDG2021+ list ($1\rightarrow$ 2 decays)}    & \multicolumn{1}{c}{$0.5486 \pm 0.0001$}  & \multicolumn{1}{c}{$0.9386 \pm 0.0003$} & \multicolumn{1}{c}{$1.4909 \pm 0.0008$} \\ \hline
        \multicolumn{1}{l}{Experiment \cite{ALICE:2019hno} }      & \multicolumn{1}{c}{$0.56965\pm 0.02505$}  & \multicolumn{1}{c}{$0.91955\pm 0.01357$} & \multicolumn{1}{c}{$1.44080\pm 0.02341$} \\ 
        \toprule
    \end{tabular}
    \caption{Mean transverse-momentum of $\pi^\pm$, $K^\pm$, and $p(\overline{p})$ produced using the blast-wave model with the default SMASH and PDG2021+ hadron lists at $5.02$ TeV Pb--Pb collisions and 0–10$\%$ centrality.
    }
    \label{tab:mean_pT}
\end{table*}
\end{widetext}

\section{Conclusions}
\label{sec:conclusions}

In this paper, we introduced the PDG2021+ hadron list, which contains up-to-date information on light and strange hadrons such as masses, widths, decay channels, and branching ratios. Using the hadron resonance gas, the PDG2021+ list was compared to lattice QCD results on partial pressures and diagonal and off-diagonal susceptibilities of conserved charges showing overall agreement in the low-temperature regime, which corresponds to the HRG regime of validity. 
In comparison to the previous PDG2016+ hadron list, the PDG2021+ displayed similar results, with differences at the percent level. The one notable exception was the $\Omega$ sector, where the new list added a new light $\Omega$ resonance that  led to a more consistent description with the lattice data points. 

Another key result of this paper is that we have updated the hadron resonance list used within the SMASH hadronic transport code. 
Thus, we have also made comparisons between the PDG2021+ and original SMASH lists.
In all cases, the PDG2021+ list outperformed the SMASH default list, indicating the need to include more hadron states. 
In order to further study the effect of increasing the number of resonances, we used a thermal model to fit the yields of light and strange particles to the values observed in Pb--Pb collisions at 5.02 TeV.
The extracted temperatures and chemical potentials are consistent with previous studies \cite{Alba:2020jir} and the $\chi^2$ fit quality is improved when more hadrons are considered, even with respect to the PDG2016+ list. 

A common limitation found in transport approaches is that the used collision criterion only allows for $1\to2$-body decays. We used a blast-wave model to compare the transverse-momentum spectra of identified particles in the cases where multi-body decays are treated in full and where intermediate states with $1\to2$ processes are used. We found a change in the slope of the pion spectra, which could have an effect when bulk viscosity is extracted from hydrodynamic simulations \cite{Ryu:2015vwa}. 
More importantly, we also explored the scenario where the blast-wave model was coupled to SMASH and calculated the spectra and $\langle p_T\rangle$. We found a slight sensitivity to the number of states in the particle list and a $\sim 3\%$ overestimation of the pion mean transverse momentum stemming from the conversion from multi-particle decays into $1\to2$ body decay chains.

As we move forward, it is imperative to integrate our findings with hydrodynamics through the equation of state and in hadronic afterburners through the available decay channels. With the current pipeline now established, the addition of more states becomes a streamlined process, paving the way for future advancements when more states become experimentally accessible once the K0L \cite{KLF:2020gai} and COMPASS++/AMBER \cite{Ketzer:2019wmd,Wallner:2022scd} collaborations begin taking data. 

\section{Acknowledgments }
This work was supported in part by the NSF within the framework of the MUSES collaboration, under grant number OAC-2103680. This material is based upon work supported by the
National Science Foundation under grants No. PHY-2208724 and No. PHY-2116686 and in part by the U.S. Department of Energy, Office of Science, Office of Nuclear Physics, under Award Number DE-SC0022023. J.S.S.M. acknowledges support from Consejo Nacional de Ciencia y Tecnolog\'ia (CONACYT) under SNI Fellowship I1200/16/2020 and Agencia Nacional de Investigaci\'on y Desarrollo (ANID) BECAS/DOCTORADO BECAS CHILE 72240405. J.N.H. acknowledges the support from the US-DOE Nuclear Science Grant No. DE-SC0020633. The authors also acknowledge support from the Illinois Campus Cluster, a computing resource that is operated by the Illinois Campus Cluster Program (ICCP) in conjunction with the National Center for Supercomputing Applications (NCSA), and which is supported by funds from the University of Illinois at Urbana-Champaign. R.H. is supported by the Helmholtz Forschungsakademie Hessen für FAIR (HFHF), and by the Deutsche Forschungsgemeinschaft (DFG, German Research Foundation) – Project number 315477589 – TRR 211. J.H. acknowledges the support by the DFG SinoGerman project (project number 410922684). J.M.K. is supported by an Ascending Postdoctoral Scholar Fellowship from the National Science Foundation under Award No. 2138063.

\bibliography{main}
\pagebreak
\appendix
\section{Hadron spectrum}
\label{appendix:hadron_spectrum}
The statistical bootstrap model is an approach that aims at characterizing nuclear matter, where the interactions among hadrons are thought to be well approximated by the formation of more massive hadron resonances, leading to a description of a gas of non-interacting heavy states. 
The information on the composition of and how rapidly these heavier states decay is contained in the density of states $\rho(m)$ that arises as the solution to the bootstrap condition,
\begin{equation}\label{eq:bootstrap_condition}
\begin{split}
    \rho(m, V_0) =\ &\delta(m-m_0)+\sum_N\frac{1}{N!}\left[ \frac{V_0}{(2\pi)^3}\right]^{N-1} \\
    & \times\int\prod_{i=1}^N\left[ dm_i\,\rho(m_i)d^3p_i\right]\delta^4\left(\sum_i p_i-p \right),
\end{split}
\end{equation}
where $m_0$ is the mass of the lightest hadron of the spectrum and $V_0$ is the size of the system. 
Hagedorn realized that solutions to the self-similar 
Eq.~(\ref{eq:bootstrap_condition}) ought to have a rising exponential behavior $\sim m^{-a}e^{m/T_H}$ for some power $a$ and improved versions of the model led to solutions of the form
\begin{equation}\label{eq:bootstrap_solution}
    \rho(m) = \frac{A}{\left( m^2 + m_r^2 \right)^{5/4}} e^{m/T_H},
\end{equation}
where $A$ and $T_H$ are free parameters and $m_r=500$ MeV; this solution has been shown to give lower values of the temperature $T_H$, closer to the transition temperature from lattice QCD, and approaches Hagedorn's initial \textit{ansatz} asymptotically in the $m\rightarrow \infty$ limit \cite{Hagedorn:1965st, Hagedorn:1968jf, Frautschi:1971ij, Nahm:1972zc, Broniowski:2004yh, Lo:2015cca}.

In order to compare to experimental data, it is actually more useful to compute the cumulative number of states of mass lower than $m$, given by
\begin{equation}\label{eq:number_of_states_theory}
    N_\text{theory}(m) = \int_{0}^{m} \rho(m')\, dm',
\end{equation}
while the total number of experimentally measured states is obtained via
\begin{equation}\label{eq:number_of_states_experiment}
    N_\text{exp.}(m) = \sum_{i\in\text{hadr.\\ list}} d_i\, \Theta(m-m_i),
\end{equation}
where $d_i$ is the degeneracy of the state $i$ and $\Theta(m)$ is the Heaviside function.

\begin{figure}[!ht]
    \includegraphics[width=0.43\textwidth]{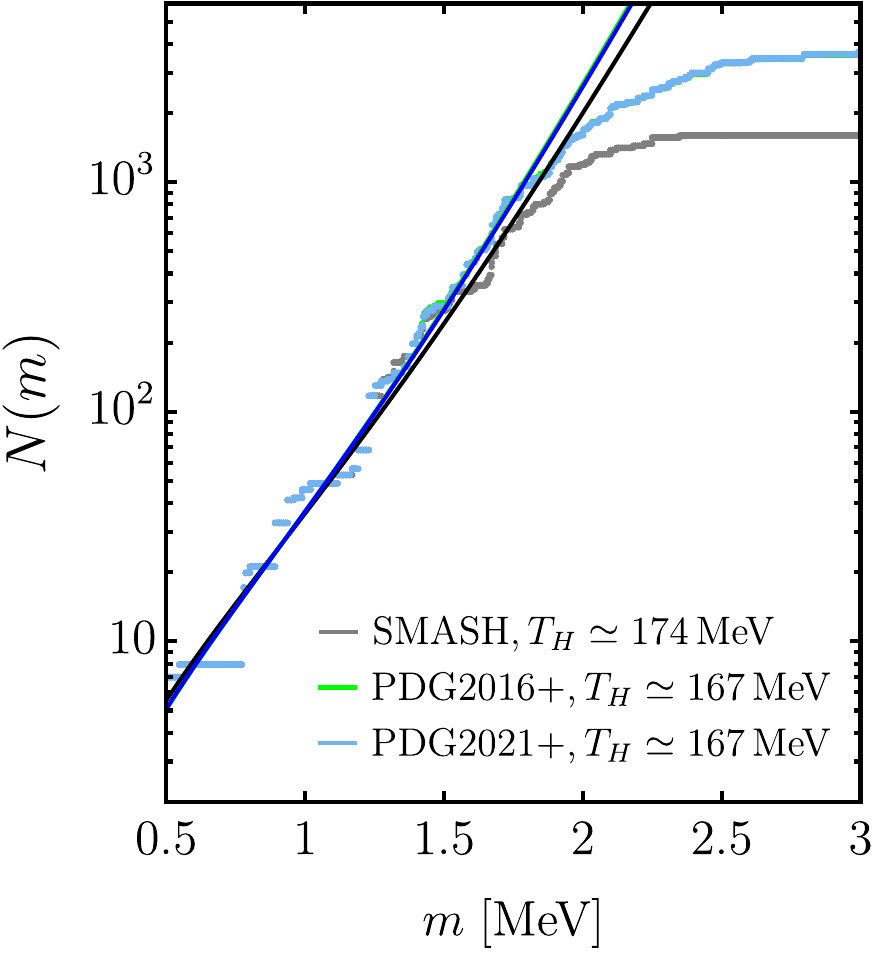}
    \centering
    \caption{Spectra of hadron states according to the default particle list included in SMASH (gray points) and PDG2021+ (blue points) and the corresponding Hagedorn fit (black and blue lines, respectively). The inclusion of more resonances lowers the Hagedorn temperature obtained from the fit.
    }
    \label{fig:hagedorn_spectrum}
\end{figure}

Figure~\ref{fig:hagedorn_spectrum} shows the number of experimentally observed states up to a given mass, $N_\text{exp.}(m)$, and the fit functions $N_\text{theory}(m)$ corresponding to the SMASH and PDG2021+ hadron lists.
The values of $A$ and $T_H$ were found performing a least-squares fit of \cref{eq:number_of_states_theory} to $N_\text{exp.}$ for each of the particle lists.
To preserve the quality of the fit, only resonances up to a mass of $1.9$ GeV were taken into account.
The resulting fit parameters were $A=0.5575$ GeV$^{3/2}$ and $T_H=174.2$ MeV for the SMASH hadron list and $A=0.4735$ GeV$^{3/2}$ and $T_H=167.2$ MeV in the case of the PDG2021+.

The inclusion of additional resonances has the effect of making the spectrum more steep, thus lowering the limiting temperature, bringing it even closer to lattice results \cite{Bazavov:2011nk, Bhattacharya:2014ara, Bellwied:2015rza, Bonati:2015bha, Bonati:2018nut, HotQCD:2018pds, Borsanyi:2020fev}, in agreement with previous studies \cite{Noronha-Hostler:2016ghw}. 
Hence, the persisting difference between the Hagedorn temperature with an extended spectrum and the pseudo-critical temperature serves as an indication that more hadron states are still to be observed in experiments.

\section{Hadron resonance gas with PDG2021+} 
\label{sec:HRG_with_PDG2021Plus}

Although the above expressions in \cref{eq:number_densities,eq:particle_multiplicity} depend on the three chemical potentials associated with conserved charges, $\mu_B$, $\mu_S$, and $\mu_Q$, these are not completely independent. 
In heavy-ion collisions, the nature of the colliding nuclei and the observation of net-strange neutrality permit to set constraints on the values of the chemical potentials via
\begin{align}\label{eq:net_charge_conditions}
    n_Q^\text{net}(T,\mu_B,\mu_S,\mu_Q)&=x\, n_B^\text{net}(T,\mu_B,\mu_S,\mu_Q),\\
    n_S^\text{net}(T,\mu_B,\mu_S,\mu_Q)&=0,
\end{align}
where the net densities of baryon number $n_B^\text{net}$, electric charge $n_Q^\text{net}$, and strangeness $n_S^\text{net}$ are calculated using
\begin{equation}\label{eq:net_number_densities}
    n_q^\text{net}(T,\mu_B,\mu_S,\mu_Q) = \sum_i X_i^q\ n_i(T, \mu_B,\mu_S,\mu_Q),
\end{equation}
effectively reducing the dependency of the number densities to the baryon chemical potential $\mu_B$ and the temperature $T$. 
The numerical factor $x$ in \cref{eq:net_charge_conditions} is obtained from the ratio of protons to baryons, namely, $x\simeq0.4$ for Au$+$Au and Pb$+$Pb collisions. 

\subsection{Comparisons to lattice QCD}
\subsubsection{Partial pressures}
\label{subsubsec:partial_pressures}
Within the HRG model, the pressure at finite chemical potential can be obtained by integrating \cref{eq:individual_pressure}, computed by expanding the logarithm and integrating term-by-term to obtain a series expansion containing the modified Bessel function of the second kind:
\begin{align}
    \frac{p_{B_iS_iQ_i}}{T^4} =&\  \frac{d_i}{2\pi^2}\left( \frac{m_i}{T} \right)^2 \exp\left(B_i\hmu_B+S_i\hmu_S+Q_i\hmu_Q\right) \nonumber \\ &\times \sum_{N=1}^\infty \frac{\left[ (-1)^{B_i+1}\right]^{N+1}}{N^2} K_2\left( N\frac{m_i}{T} \right).
\end{align}
Because we are interested in the regime of temperatures $T\leq170$ MeV and the masses of the lightest baryons are $m_N \simeq 1$ GeV, we can approximate the above expression for large values of the ratio $m_i/T$. 
In this approximation, we take only the first term in the series above, due to the Bessel function becoming suppressed at higher orders as $\sim \exp\left( -N m_i/T \right)$. 
Of course, this argument remains formally valid for all but the lightest mesons; in practice, however, the approximation can also hold for pions up to a few-percent deviation.
Hence, the pressure for particle $i$ takes the form

\begin{equation}
    \frac{p_{B_iS_iQ_i}}{T^4} = \frac{d_i}{2}\, \phi(T, m_i) \exp\left(B_i\hmu_B+S_i\hmu_S+Q_i\hmu_Q\right),
\end{equation}
where
\begin{equation}
    \phi(T,m_i) = \frac{1}{\pi^2}\left( \frac{m_i}{T} \right)^2 K_2\left( \frac{m_i}{T} \right).
\end{equation}
\noindent
Indeed, the next-to-leading terms can be identified as quantum corrections and this constitutes the Boltzmann approximation.

In order to compare the various contributions to the total pressure coming from hadrons that possess the same quantum numbers, we sum over all particles that have the same $\{ B,S,Q \}$ charges, 

\begin{align}
    \frac{p_{BSQ}}{T^4} &= \sum_{i} \frac{p_{B_iS_iQ_i}}{T^4} \delta_{BB_i}\delta_{SS_i}\delta_{QQ_i} \nonumber \\
    &= \exp\left(B\hmu_B+S\hmu_S+Q\hmu_Q\right)\, \phi_{BSQ}(T),
\end{align}
where $\phi_{BSQ}$ is defined by
\begin{equation}
    \phi_{BSQ}(T) = \sum_{i} d_i\, \phi(T, m_i)\, \delta_{BB_i}\delta_{SS_i}\delta_{QQ_i}.
\end{equation}
With the explicit contribution from antiparticles and setting $\hmu_Q=0$ for simplicity, the total pressure can then be written as 

\begin{align}
    \frac{p}{T^4} &= \tilde{\phi}_{0}(T) + \sum_{B,S} \tilde{\phi}_{BS}(T)\, \cosh(B\hmu_B+S\hmu_S) \nonumber \\
    &= \tilde{\phi}_{0}(T) 
    + \tilde{\phi}_{0|1|}(T)\cosh(\hmu_S) \nonumber \\
    & \hspace{39pt} + \tilde{\phi}_{10}(T)\cosh(\hmu_B) \nonumber \\
    & \hspace{39pt} + \tilde{\phi}_{1|1|}(T)\cosh(\hmu_B-\hmu_S) \nonumber \\
    & \hspace{39pt} + \tilde{\phi}_{1|2|}(T)\cosh(\hmu_B-2\hmu_S) \nonumber \\
    & \hspace{39pt} + \tilde{\phi}_{1|3|}(T)\cosh(\hmu_B-3\hmu_S), \nonumber
\end{align}
where each term in the sum corresponds to the partial pressure associated with a particular set of quantum numbers. Notice that the dimensionless pressure coefficients $\tilde{\phi}_{BS}$ are now linear combinations of the more general $\phi_{BSQ}$; for instance, the kaon contribution is $\tilde{\phi}_{0|1|}=\phi_{0|1|0}+\phi_{0|1||1|}$.
For further details see \cite{Bazavov:2013dta, Noronha-Hostler:2016rpd, Alba:2017mqu}.

In order to scrutinize further the updates to the hadron spectrum with the PDG2021+ list, we show its agreement with the lattice data in the different sectors along with the results from HRG model calculations with older hadron lists.
We compare the results from the new list with the list that first incorporated additional PDG states beyond those that are well-established, known as PDG2016+, as described in detail in Refs. \cite{Alba:2017mqu,Alba:2020jir}.
In addition, the PDG2021+ list has been developed for use in the simulation of strongly-interacting matter, and as such, we compare it to the hadron list from the SMASH hadronic transport framework.

\begin{figure}[tb]
    \includegraphics[width=0.94\linewidth]{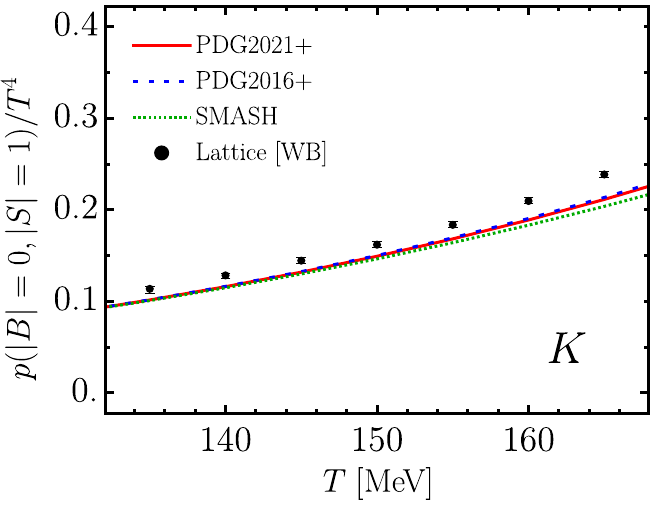}
    \centering
    \caption{Contribution to the total pressure from $K$ resonances within the ideal hadron resonance gas using the PDG2021+ (solid red) particle list, compared to the lattice results from \cite{Alba:2017mqu} and to the PDG2016+ (dashed blue) and SMASH (dotted green) hadron lists.}
    \label{fig:kaon_partial_pressures}
\end{figure}

\begin{figure}[t]
    \includegraphics[width=0.96\linewidth]{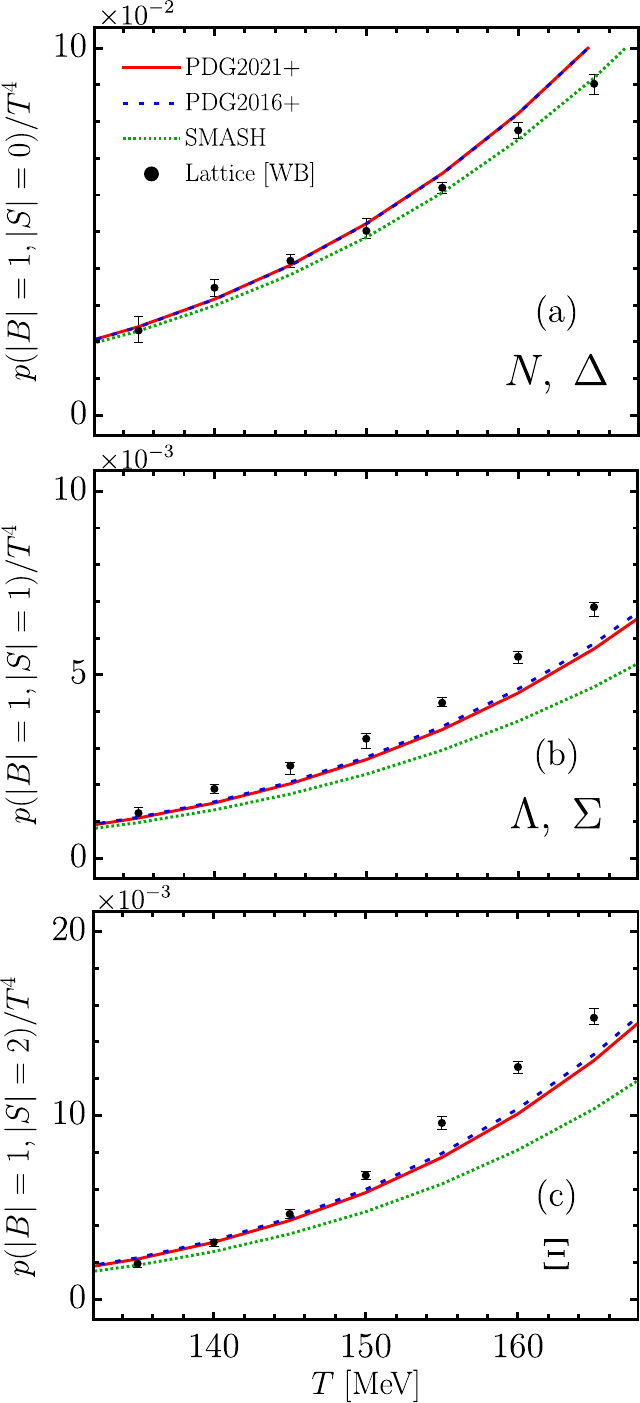}
    \centering
    \caption{Partial pressures of (a) $N$ and $\Delta$, (b) $\Lambda$ and $\Sigma$, (c) $\Xi$, and (d) $\Omega$ resonances, as obtained from the ideal hadron resonance gas  using the PDG2021+ (solid red), PDG2016+ (dashed blue), and SMASH (dotted green) particle lists, compared to the lattice results from \cite{Alba:2017mqu}.
    }
    \label{fig:baryon_partial_pressures}
\end{figure}

A comparison of the partial pressures for the PDG2021+ list with the SMASH and PDG2016+ lists is shown in \cref{fig:kaon_partial_pressures,fig:baryon_partial_pressures}. 
From this we see that the strange meson content is very similar among the three hadron lists, which is due to the fact that there have not been modern experimental facilities available to study the strange meson spectrum\footnote{The proposal from the COMPASS++/AMBER collaboration of upgrading and operating the existing facilities at the M2 beam line at CERN SPS has been approved and is expected to improve and extend the kaon spectrum \cite{Adams:2018pwt}.}. 
However, for the non-strange baryons, shown in the top left panel of Fig. \ref{fig:baryon_partial_pressures}, the SMASH list lies below the two extended lists from 2016 and 2021, while the extended lists remain in agreement. This shows that there are resonances missing in the SMASH list already in the non-strange baryon family. 
Indeed, this trend continues for the $S=1$ and $S=2$ baryon sectors, also shown in panels (b) and (c) of Fig. \ref{fig:baryon_partial_pressures}. 
On the other hand, the partial pressures in the singly and doubly strange baryon sectors continue to agree for the extended lists, PDG2016+ and PDG2021+, with deviations only at the percent level.
It it useful to note that the deviations seen in these plots between the HRG model and lattice QCD near and above the pseudo-critical temperature, $T_\text{p.c.} \sim 155$ MeV, naturally arise from the fact that the system can no longer be described with hadronic degrees of freedom.

\label{subsec:comparisons_to_LQCD}
\subsubsection{Susceptibilities}
\label{subsubsec:susceptibilities}

\begin{figure*}[!htbp]
    \includegraphics[width=0.80\textwidth]{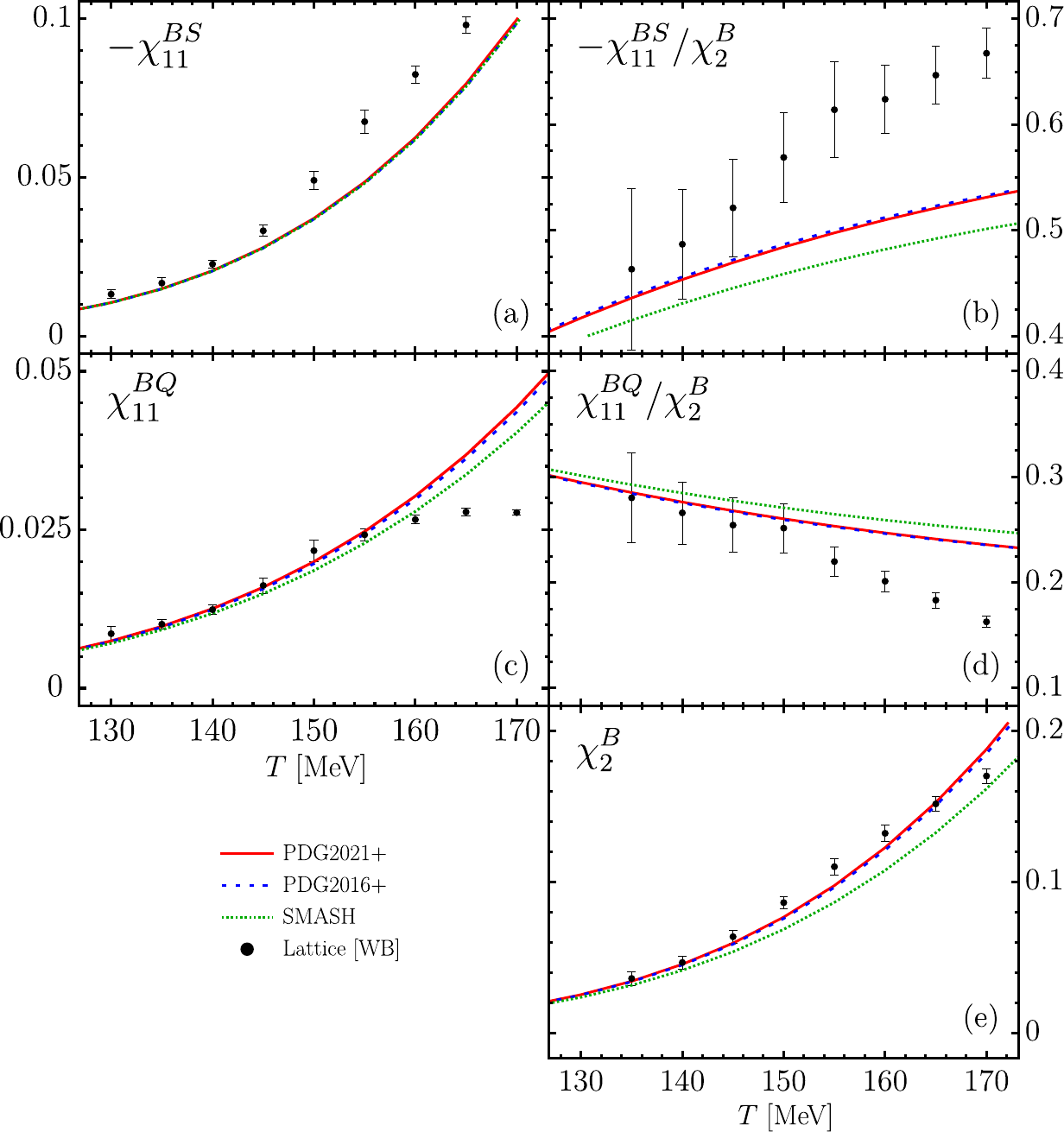}
    \centering
    \caption{Cross-charge susceptibility $\chi_{11}^{BQ}$ at vanishing chemical potential, normalized by $\chi_2^B$ as a function of temperature within the ideal HRG model using the PDG2021+ (solid red), PDG2016+ (dashed blue), and SMASH (dotted green) particle lists, compared to the lattice results from Refs. \cite{Bellwied:2015lba, Bellwied:2019pxh}.}
    \label{fig:chi_grid}
\end{figure*}

Next, we show off-diagonal ratios for baryon-charge and baryon-strangeness susceptibilities. These ratios in particular are sensitive to the effects of the hadron spectrum, as discussed in Ref. \cite{Karthein:2021cmb}.

The middle-right panel of \cref{fig:chi_grid} shows the ratio of charged baryons to all baryons, $\chi_{11}^{BQ}/\chi_2^{B}$, for the three hadron lists we compare.
As with the higher order strangeness susceptibility ratios, the augmented lists PDG2016+ and PDG2021+ agree very well. The SMASH list lies slightly above, though the difference is not dramatic.
We reiterate that, because the lists perform so similarly for these susceptibilities, their agreement with lattice results is also very similar.
In the new hadron list PDG2021+, some states that were ambiguous in the 2016 version of the PDG list were resolved and were consequently removed from the updated PDG2021+ list, as shown in Table \ref{tab:particle_changes}.
In light of this, the new PDG2021+ list is slightly above its predecessor for $\chi_{11}^{BQ}/\chi_2^{B}$, which indicates that the fraction of charged baryons is slightly larger.

At the same time, we see that the ratio of strange baryons to all baryons is slightly smaller for PDG2021+ compared to PDG2016+, as shown in the top-right panel of Fig. \ref{fig:chi_grid}.
This shows a greater separation in the SMASH list compared to the extended spectrum lists, as strange baryons contribute less to the total number of baryons for this list.

\begin{figure}[!htbp]
    \includegraphics[width=0.95\linewidth]{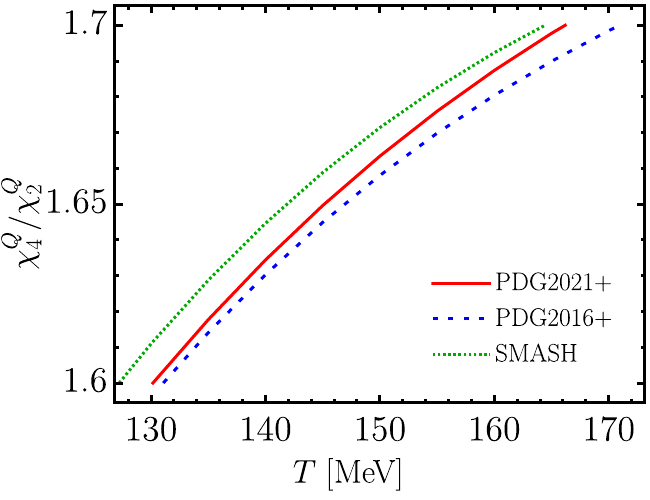}
    \centering
    \caption{Electric charge susceptibility $\chi_4^Q$ at vanishing chemical potential, normalized by $\chi_2^Q$ as a function of temperature within the ideal HRG model using the PDG2021+ (solid red), PDG2016+ (dashed blue), and SMASH (dotted green) particle lists. 
}
    \label{fig:ratio_chiQ4_to_chiQ2}
\end{figure}

Finally, we compare the lists for the electric charge susceptibility ratio $\chi_4^Q/\chi_2^Q$. Continuum extrapolated results from lattice QCD are not available at the moment, because of the difficulty in obtaining a continuum limit for $\chi_4^Q$. Because all hadrons except the $\Delta$ baryons have unit (or zero) charge, this ratio can deviate from 1 only due to these isospin multiplets, for each of which there is a $\left|Q\right|=2$ state. Every other singly charged state will bring the ratio closer to 1, including heavier (strange) resonances. This explains why the SMASH curve is above the PDG ones, as it contains fewer states in general, while the number of $\Delta$ baryons is similar. The PDG2021+, on the other hand, falls above the PDG2016+ because of the recent decrease in the mass of a couple of the lightest $\Delta$ resonances, as can be seen in the top right panel of Fig.~\ref{fig:PDG21Plus_vs_PDG16Plus_vs_SMASH_spectra_comparison}.

\subsection{Thermal Models}
\label{appendix:thermal_models}

\begin{figure}[!tb]
    \includegraphics[width=0.48\textwidth]{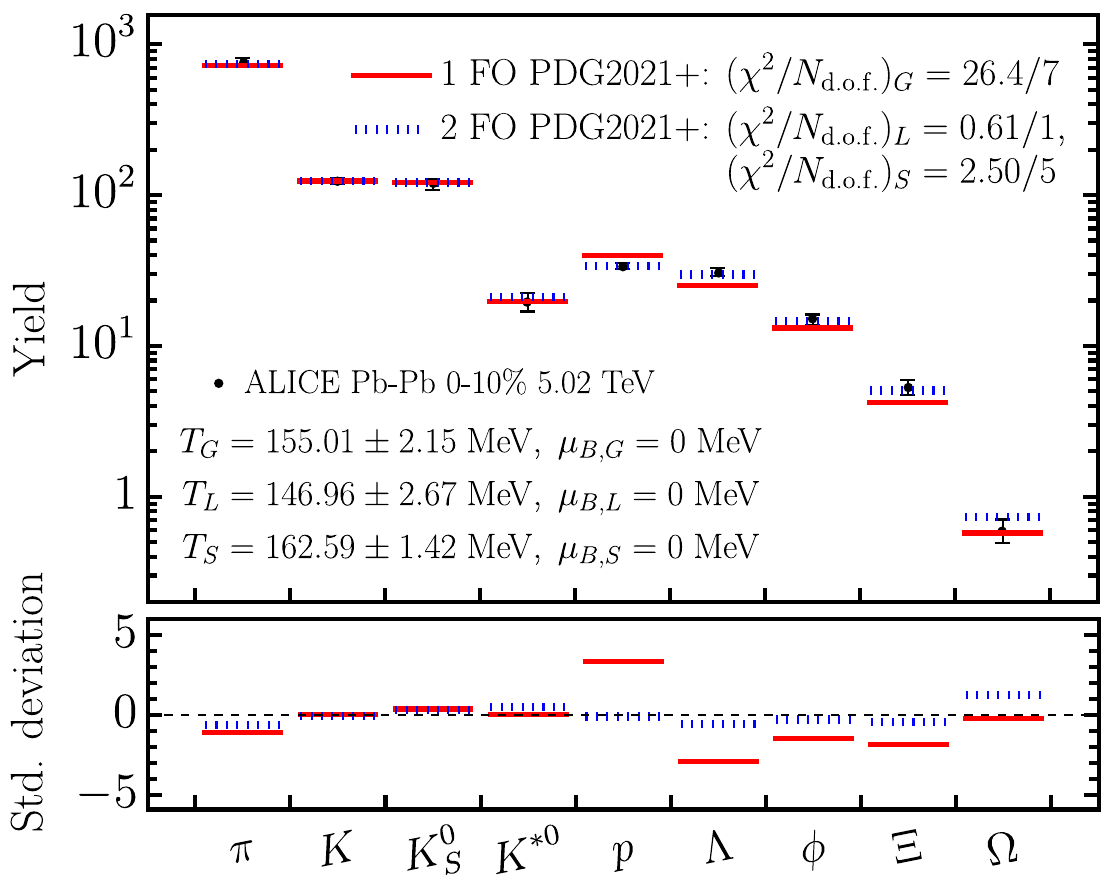}
    \centering
    \caption{(Solid) Single freeze-out temperature particle yield fit using the PDG2021+ resonance list with extracted volume $V_G=5287.69\pm639.26$ fm$^3$ and (dotted) Two-freeze-out temperature particle yield fit using the PDG2021+ resonance list with $V_L=7717.14\pm1139.73$ fm$^3$ and $V_S=3753.62\pm283.55$ fm$^3$ as the extracted volumes. Subscripts $G$, $L$, and $S$ stand for ``Global'', ``Light'', and ``Strange'' particles, respectively. Experimental data points correspond to ALICE Pb--Pb $0$–$10\%$ collisions at $5.02$ TeV \cite{ALICE:2019hno,ALICE:2021ptz,Bhasin:2022rpo}. 
    \label{fig:yield_ALICE_PDG21Plus}}
\end{figure}

\begin{figure}[!b]
    \includegraphics[width=0.48\textwidth]{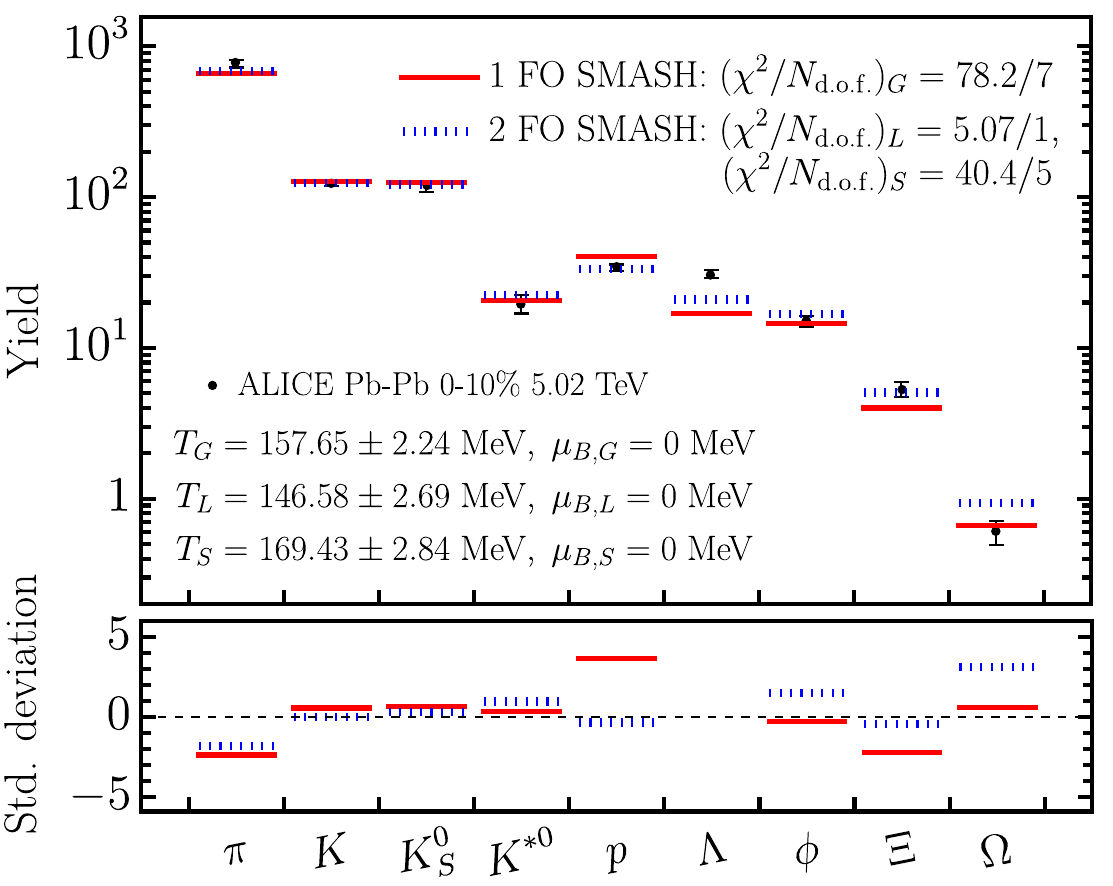}
    \centering
    \caption{(Solid) Single freeze-out temperature particle yield fit using the SMASH resonance list with extracted volume $V_G=5075.0\pm603.7$ fm$^3$ and (dotted) Two-freeze-out temperature particle yield fit using the SMASH resonance list with $V_L=8631.94\pm1239.49$ fm$^3$ and $V_S=3030.02\pm417.28$ fm$^3$ as the extracted volumes. Subscripts $G$, $L$, and $S$ stand for ``Global'', ``Light'', and ``Strange'' particles, respectively. Experimental data points correspond to ALICE Pb--Pb $0$–$10\%$ collisions at $5.02$ TeV \cite{ALICE:2019hno,ALICE:2021ptz,Bhasin:2022rpo}.
    \label{fig:yield_ALICE_SMASH}}
\end{figure}

Figure~\ref{fig:yield_ALICE_PDG21Plus} illustrates the total yields for all stable particles in the single and two-freeze-out scenarios using the PDG2021+ resonance list with full decays, as well as the corresponding standard deviations from experimental data for central ALICE Pb--Pb $0$–$10\%$ events at $5.02$ TeV, along with the extracted fit parameters and their associated uncertainties \cite{ALICE:2019hno,ALICE:2021ptz,Bhasin:2022rpo}. 
In a similar manner, \cref{fig:yield_ALICE_SMASH} displays the total yield when employing the SMASH hadron list. 

The best-fit parameters for all thermal fits are summarized in \cref{tab:fit_summary}, where a consistent improvement in the quality of the fit is observed when using the PDG2021+ particle list with both decay chains and full decays, as indicated by the value of $\chi^2/N_\text{d.o.f.}$. In the case of the two freeze-out scenario, we have computed the total entropy for light and strange particles, for both the SMASH and PDG2021+ hadron lists, as summarized in \cref{tab:fit_entropies}. As discussed in Refs.~\cite{Bugaev:2013sfa,Bugaev:2016voh}, the two freeze-out scenarios must be connected by entropy conservation and hence, must have the same total entropy. From \cref{tab:fit_entropies}, we see that total entropy at chemical freeze-out is consistent between light and strange freeze-outs within error bars.

\begin{table*}[!htbp]
    \begin{tabular}{lccc}
        \toprule
        Particle list & \multicolumn{1}{c}{\hspace{7mm}Particle subset}\hspace{7mm} &
            \multicolumn{1}{c}{\hspace{5mm}$T_\text{ch}$ [MeV]}\hspace{5mm}     &
            \multicolumn{1}{c}{\hspace{5mm}$s_\text{ch}V_\text{ch}$}\hspace{5mm} \\ \hline
        \multirow{3}{*}{SMASH list} 
        & \multicolumn{1}{c}{All}  
        & \multicolumn{1}{c}{$157.65\pm 2.24$} 
        & \multicolumn{1}{c}{$12846.04\pm1528.11$} \\ 
        & \multicolumn{1}{c}{Light}  
        & \multicolumn{1}{c}{$146.58\pm 2.69$} 
        & \multicolumn{1}{c}{$13755.00\pm1975.13$} \\
        & \multicolumn{1}{c}{Strange}  
        & \multicolumn{1}{c}{$169.43\pm 2.84$} 
        & \multicolumn{1}{c}{$12307.43\pm1694.92$} \\
        \hline
        \multirow{3}{*}{\shortstack{PDG2021+ list\\ ($1 \to 2$ decays)}}
        & \multicolumn{1}{c}{All}  
        & \multicolumn{1}{c}{$154.80\pm 2.22$} 
        & \multicolumn{1}{c}{$14852.69\pm1863.55$} \\ 
        & \multicolumn{1}{c}{Light}  
        & \multicolumn{1}{c}{$145.57\pm 2.66$} 
        & \multicolumn{1}{c}{$16426.66\pm2417.21$} \\ 
        & \multicolumn{1}{c}{Strange}  
        & \multicolumn{1}{c}{$163.04\pm2.13$} 
        & \multicolumn{1}{c}{$14066.64\pm1571.91$} \\ 
        \hline
        \multirow{3}{*}{\shortstack{PDG2021+ list\\ ($1 \to \text{all}$ decays)}}
        & \multicolumn{1}{c}{All}  
        & \multicolumn{1}{c}{$155.01\pm2.15$} 
        & \multicolumn{1}{c}{$15152.93\pm1831.93$}\\ 
        & \multicolumn{1}{c}{Light}  
        & \multicolumn{1}{c}{$146.96\pm2.67$} 
        & \multicolumn{1}{c}{$15946.85\pm2355.16$} \\ 
        & \multicolumn{1}{c}{Strange}  
        & \multicolumn{1}{c}{$162.59\pm1.42$} 
        & \multicolumn{1}{c}{$14583.00\pm1101.61$} \\ 
        \toprule
    \end{tabular}
    \caption{Temperature and total entropy obtained at chemical freeze-out obtained from thermal yield fits (see \cref{fig:yield_ALICE_PDG21Plus,fig:yield_ALICE_SMASH}) performed for the SMASH and PDG2021+ hadron lists. The ``All'' results correspond to the single freeze-out scenario, where all particles are fitted simultaneously while the ``Light'' and ``Strange'' results correspond to the two freeze-out scenario, in which a separate fit is performed for both sets of particles (see main text for details). Experimental data points correspond to ALICE Pb--Pb $0$–$10\%$ collisions at $5.02$ TeV \cite{ALICE:2019hno,ALICE:2021ptz,Bhasin:2022rpo}. }
    \label{tab:fit_entropies}
\end{table*}

\section{Particle identification numbering scheme for new particles}
\label{appendix:PIDs}
The Monte Carlo Particle Numbering Scheme is commonly used to identify particles in a standardized manner across event generators, detector simulations, and analysis packages \cite{ParticleDataGroup:2022pth}.
The latest version of the scheme was adopted in 1998 and although it was designed to be revised and updated, it has a few pitfalls. 
It consists of a string of 7 digits and an overall sign:
\begin{equation}
    \pm n n_r n_L n_{q_1} n_{q_2} n_{q_3} n_J, \nonumber
\end{equation}
whose content meaning varies depending if the particle in question is a meson or a baryon.
The last digit, $n_J=2J+1$ gives the particle's spin but does not cover particles of spin $J>4$, of which the PDG2021+ has 40 instances.
In general, the digits $n_{q_{1-3}}$ are used to specify the quark content of the hadron. 
However, in mesons where the quark content is unknown or not well defined, it is fixed to $n_{q_1} n_{q_2} n_{q_3}=033$. 
Furthermore, there are baryon species that have the same quark content, such as $N$ and $\Delta$, and $\Lambda$ and $\Sigma$ baryons. 
In these cases, the ordering of these digits is such that the lightest particle between those with the same $n_{q_1} n_{q_2} n_{q_3} n_J$ keeps the smaller number.
The digit $n_r$ labels radially excited mesons above the ground state and for baryons it is intended to be assigned following the harmonic oscillator model, although in practice is only implemented for the heaviest $\Xi$ and $\Omega$ resonances.
For mesons, the digit $n_L$ normally distinguishes states with different orbital angular momentum but for states that are not well established, it increases with mass.
On the other hand, for baryons, $n_L$ lifts any degeneracy among particles with the same $n_r$ and $n_J$.
Finally $n=0$ is set for all established mesons and $n=9$ otherwise.
Unconfirmed or established baryons with incomplete information are suggested to be prepended with $nn_r=99$.

In order to be consistent when naming new hadrons, and to encode useful information on the particle identification number, we adopted a few additional conventions:
\\~\\
\noindent
\textit{Mesons}
\begin{itemize}
    \item Treat $n_L n_{q_1} n_{q_2} n_{q_3} n_J$ following the standard.
    \item Use $n=7$ instead of $n=9$ when conflicts occur.
    \item Use $n=8$ when $J > 4$ and set $n_J = 2J + 1 - 10$.
\end{itemize}
\noindent
\textit{Baryons}
\begin{itemize}
    \item Treat $n_L n_{q_1} n_{q_2} n_{q_3} n_J$ following the standard.
    \item Use $nn_r = 98$ instead of $nn_r = 99$.
    \item Use $nn_r = 99$ when $J \geq 11/2$ and set $n_J = 2J + 1 - 10$.
    \item Use $nn_r = 99$ when $J = 9/2$ and set $n_J = 8$.
\end{itemize}

\section{Well-known hadron states}
\label{appendix:PDG2021}
\begin{figure}[!hb]
    \includegraphics[width=0.48\textwidth]{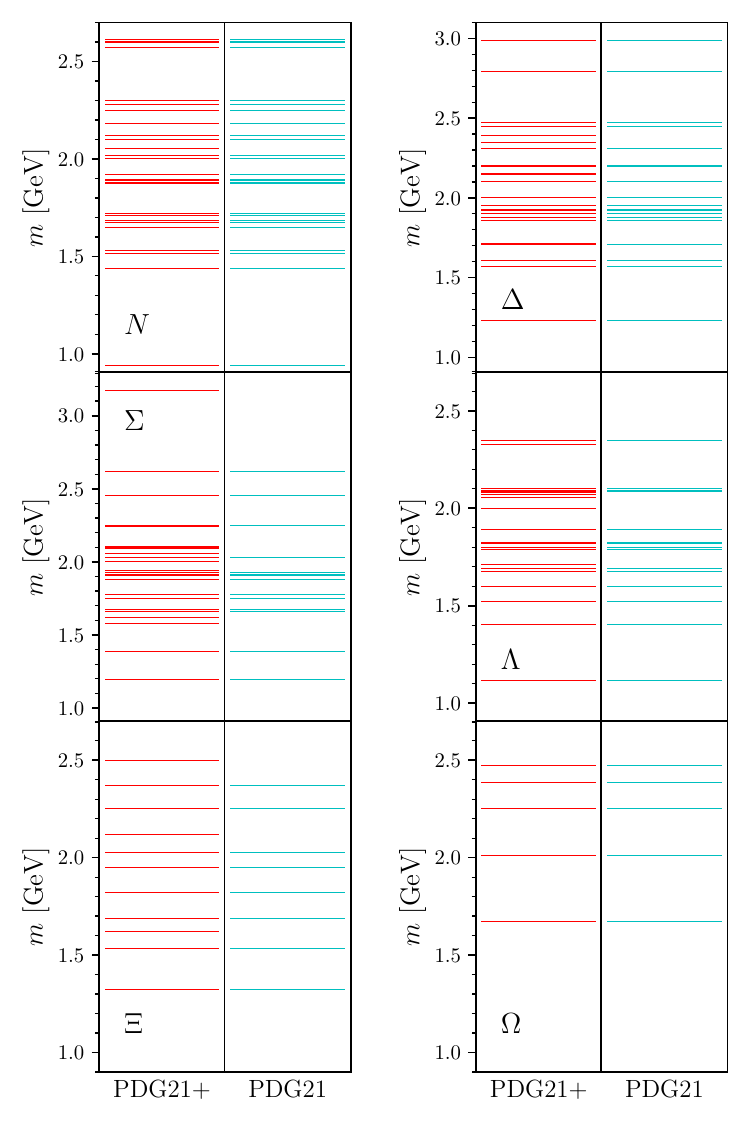}
    \centering
    \caption{Comparison of baryon resonances mass spectra per family between the PDG2021+ (red) and PDG2021 (cyan) hadron lists. The PDG2021+, is an extended list that contains all hadrons (up to strange in flavor) that have been experimentally observed. The PDG2021 is a subset that includes only the most well-established states for which there is sufficient information about their existence and properties. 
    }
    \label{fig:PDG21Plus_vs_PDG21_spectra_comparison}
\end{figure}

For applications where only the most up-to-date and best experimentally known states are needed, the PDG2021 list is also provided, which restricts to only (****) and (***) states.
In \cref{fig:PDG21Plus_vs_PDG21_spectra_comparison}, we show the comparison between the well-established states, labeled as PDG2021 and the extended PDG2021+, which in addition to all the particles included in the PDG2021, also includes those states that have been observed but require further experimental confirmation. 

\section{Adaptation of PDG2021+ into SMASH}
\label{appendix:cross-section_details}
\subsection{Intermediate states}
In total, there are 4197 decay channels included in the PDG2021+, of which 3637 are $1\rightarrow2$-body decays and the rest are 3-or-4-product reactions. In order to be incorporated into SMASH, multi-body decays need to be modeled through decay chains using intermediate states. For any given decay chain, intermediate states have to possess the appropriate symmetries that lead to electric charge, parity, and isospin conservation. Additionally, energy and momentum are conserved at each step of the chain, so a decay is allowed only if the sum of the mass of the daughter particles is less than that of the mother.

Depending on the decaying particle, several candidate intermediate states may arise; here we have prioritized for the intermediate states that have the lowest total combined mass. Light resonances with multi-particle decay channels can have instances where all possible intermediate states have a total combined mass greater than the mother particle. In such cases, the decay channel in question is omitted and the remaining branching ratios are normalized.

\subsection{Cross-sections}
SMASH computes the total cross-section in the low $\sqrt{s}$ region in a ``bottom-up'' approach, where the partial resonant contributions are added. For most measured elementary interactions, the default SMASH hadron list leads to a good agreement with experimental cross-sections, both inclusive and exclusive.
For meson-meson and meson-baryon processes, the elastic cross-section is fully determined by resonance excitation and decay, while for baryon-baryon it is parametrized. 
In the high $\sqrt{s}$ region, both elastic and total cross-sections are parametrized, such that the inelastic part ($\sigma_\mathrm{tot}-\sigma_\mathrm{el}$) is handled by PYTHIA 8  via string fragmentation \cite{Sjostrand:2007gs}. 
Interactions between resonances cannot be experimentally measured, for them the Additive Quark Model (AQM) \cite{Sorge:1990fw} is assumed in the high energy regime.

\begin{figure}[t]
    \includegraphics[width=0.48\textwidth]{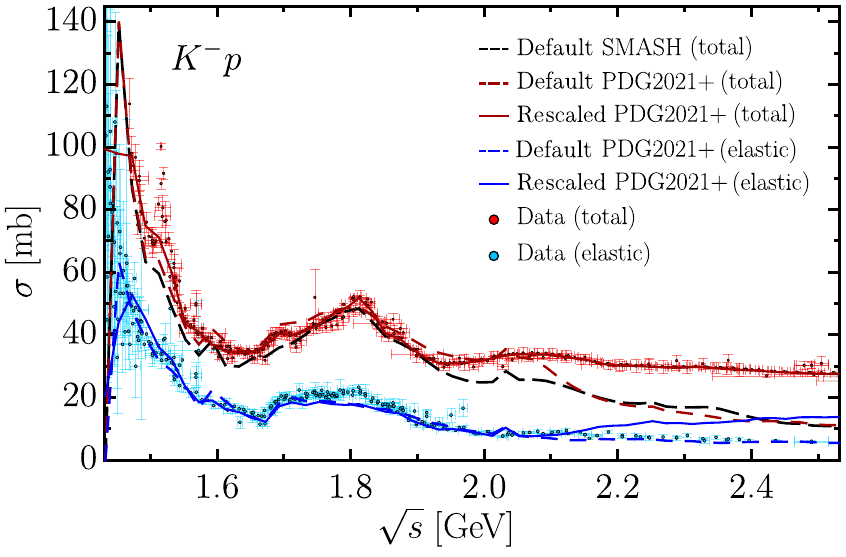}
    \centering
    \caption{Cross-sections of kaon-proton interactions by outgoing particles, using the PDG2021+ hadron list. The total (elastic) cross-section obtained with the PDG2021+ list is shown in red (blue), while the total cross-section with the default SMASH hadron list is shown in black. Dashed lines are the default bottom-up implementation of SMASH, and solid lines show the re-scaled cross-sections (see text). Points are world data.}
    \label{fig:xs_Kp_all}
\end{figure}

\begin{figure}[t]
    \includegraphics[width=0.48\textwidth]{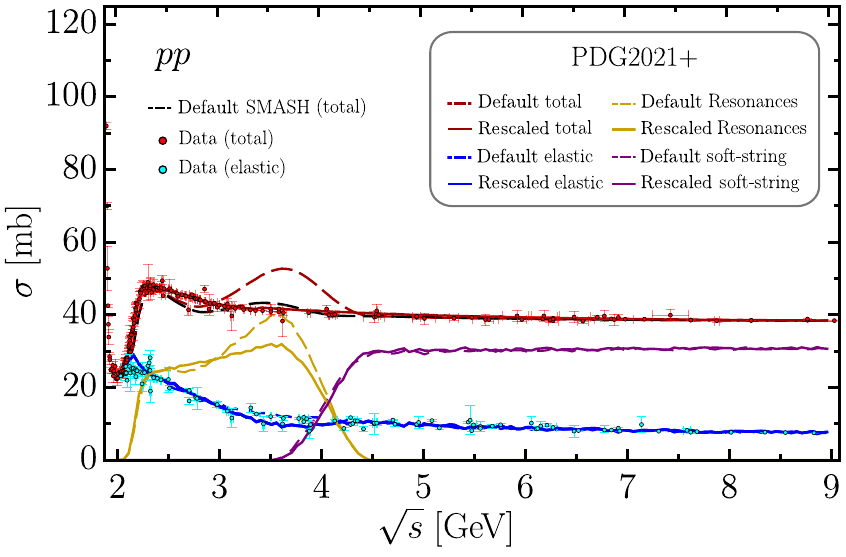}
    \centering
    \caption{Cross-sections of proton-proton interactions by outgoing particles, using the PDG2021+ hadron list. The total (elastic) cross-section obtained with the PDG2021+ list is shown in red (blue), while the total cross-section with the default SMASH hadron list is shown in black. The contribution from resonance production in $2\to2$ processes is shown in yellow, and the soft-strings handled by PYTHIA 8 are shown in purple. Dashed lines are the default bottom-up implementation of SMASH, and solid lines show the re-scaled cross-sections (see text). Points are world data.
    }\label{fig:xs_pp_all}
\end{figure}

The effect of the PDG2021+ hadron list on SMASH cross-sections is exemplified in the red curves of Figs. \ref{fig:xs_Kp_all} and \ref{fig:xs_pp_all}. 
In comparison to the dashed black lines, computed with the default SMASH list, the updated list provides a better agreement to experimental data in the strange sector (dashed red line in Fig. \ref{fig:xs_Kp_all}). This is consistent with the previous lattice QCD observations \cite{Alba:2017mqu} where the addition of one-star states considerably improved the agreement to lattice results. However, the lack of processes after $\sqrt{s}\approx2.1\ \mathrm{GeV}$ may be a further indication of missing heavier strange resonances, with the caveat that mixing and interference terms between these resonances are not properly taken into account. On the other hand, the default SMASH list already matched experimental data in the non-strange sector, so the addition of new resonances breaks this agreement (dashed red line in Fig. \ref{fig:xs_pp_all}).

Since the total cross-section is used to determine possible collisions through the geometric criterion, these mismatches would lead to an undesirable excess (lack) of inelastic $pp$ ($Kp$) interactions in the afterburner. Therefore, a ``top-down'' approach is used in the resonance region for this work, in which the sum of partial contributions is re-scaled to match total cross sections where experimental data is available. The final values used are shown in the solid lines. By construction, the solid red lines lie on top of the data, but in some cases, this procedure might lead to an overshoot of the elastic branching ratio (solid blue line in Fig. \ref{fig:xs_Kp_all}). The top-down approach will become the default behavior in the upcoming SMASH-3.1. 

Another interesting aspect that could be further investigated with the present approach are thermodynamic properties of the improved hadron gas, such as transport coefficients. Equilibrium quantities like shear and bulk viscosities, conductivities, and diffusion coefficients \cite{Rose:2017bjz, Rose:2020lfc,Rose:2020sjv, Hammelmann:2023fqw} were calculated from SMASH in an infinite matter simulation (box with periodic boundary conditions) using the Green-Kubo formalism, and a direct extension with the updated PDG2021+ list would be in principle feasible. However, one of the differences introduced with this list are decays of resonances into photons which have to be treated dynamically by the transport model, so cross sections are assigned in order to preserve detailed balance. SMASH computes these automatically, but they tend to be very small. Together with the large number of such channels, this causes the produced photons to have a very long lifetime within the simulation, which has a large effect on e.g. the shear viscosity. Since the relaxation time is increased by the long-lived photons, perturbations of the off-diagonal components of the energy momentum tensor relax very slowly, which overincreases the shear viscosity. It is currently unclear how to treat the hadronic decays into photons correctly in the box calculation in thermal and chemical equilibrium, and therefore we leave the analysis of transport coefficients for future studies.

\end{document}